\documentclass[submission, Proceedings]{SciPost}

\hypersetup{
    colorlinks,
    linkcolor={red!50!black},
    citecolor={blue!50!black},
    urlcolor={blue!80!black}
}

\usepackage[bitstream-charter]{mathdesign}
\usepackage{caption}
\usepackage{subcaption}
\usepackage{amsmath}

\usepackage{floatrow}
\newfloatcommand{capbtabbox}{table}[][\FBwidth]
\DeclareSymbolFont{usualmathcal}{OMS}{cmsy}{m}{n}
\DeclareSymbolFontAlphabet{\mathcal}{usualmathcal}

\begin{document}

\begin{center}{\Large \textbf{
Lepton Flavour Universality tests and Lepton Flavour Violation searches at LHCb\\
}}\end{center}

\begin{center}
S. Celani\textsuperscript{1}
\\
on behalf of the LHCb collaboration
\\
\end{center}

\begin{center}
{\bf 1} École Polytechnique Fédérale de Lausanne, Lausanne, Switzerland
\\
sara.celani@cern.ch
\end{center}

\begin{center}
\today
\end{center}


\definecolor{palegray}{gray}{0.95}
\begin{center}
\colorbox{palegray}{
  \begin{minipage}{0.95\textwidth}
    \begin{center}
    {\it  16th International Workshop on Tau Lepton Physics (TAU2021),}\\
    {\it September 27 – October 1, 2021} \\
    \doi{10.21468/SciPostPhysProc.?}\\
    \end{center}
  \end{minipage}
}
\end{center}

\section*{Abstract}
{\bf
The Standard Model (SM) of particle physics represents our most fundamental knowledge of elementary particles and their interactions. One of the cardinal properties of the SM which has been thoroughly studied during the past years is the so-called Lepton Flavour Universality (LFU): W and Z bosons are predicted to be equally coupled to the three lepton generations. Hints for deviations from LFU have been found by the LHCb collaboration in $\mathbf{b\to s\ell\ell}$ and $\mathbf{b\to c\ell\nu}$ decays, sparking great interest and further motivating the searches for lepton flavour violating processes. A general review of these results is presented, including prospects in view of the LHC and LHCb upgrade.
}

\vspace{10pt}
\noindent\rule{\textwidth}{1pt}
\tableofcontents\thispagestyle{fancy}
\noindent\rule{\textwidth}{1pt}
\vspace{10pt}

\section{Introduction}
\label{sec:intro}
The Standard Model (SM) predicts the electroweak interactions to have the same amplitudes for all the three different lepton generations, except for phase-space differences or helicity suppression effects. This property is called Lepton Flavour Universality (LFU) and it has been experimentally verified in meson decays ~\cite{Jpsi_LFU}~\cite{K_LFU}~\cite{pi_LFU}, $\tau$ decays~\cite{PDG} and $Z$ boson~\cite{Z_LFU} decays. Nevertheless, evidence of LFU violation has been recently observed by the LHCb collaboration in $B^+\to K^+\ell^+\ell^-$ ($\ell = e,\mu$) decays~\cite{RK_21}: yet another piece in the larger set of anomalies observed in the last decade in $B$ meson decays, which are showing a consistent tension with the SM predictions (usually referred to as {\it flavour anomalies}). A review of these anomalies, measured in $b\to s\ell\ell$ decays (Section~\ref{sec:fcnc}) and in $b\to c\nu\ell$ decays (Section~\ref{sec:cc}), is given in this proceeding.

\noindent
Beyond Standard Model (BSM) theories that could explain the anomalies, e.g. leptoquarks~\cite{LQ_14,LQ_15,LQ_15_2,LQ_16} or new heavy gauge boson such as $Z'$~\cite{Zprime_14,Zprime_15,Zprime_15_2,Zprime_15_3},  generally also imply a sizeable Lepton Flavour Violation (LFV), for which evidence has been already observed in decays involving neutrinos~\cite{nu_lfv}. An overview of the searches for decays breaking the lepton flavour conservation is reported in Section~\ref{sec:LFV}. Finally in Section~\ref{sec:conclusion}, conclusions and future prospects, given the upgrade of the LHC and the LHCb detector, are presented.

\section{Lepton Flavour Universality tests}
\label{sec:LFU}
\begin{figure}[htpb]
\centering
\begin{subfigure}[b]{0.52\textwidth}
  \centering
  \includegraphics[width=\textwidth]{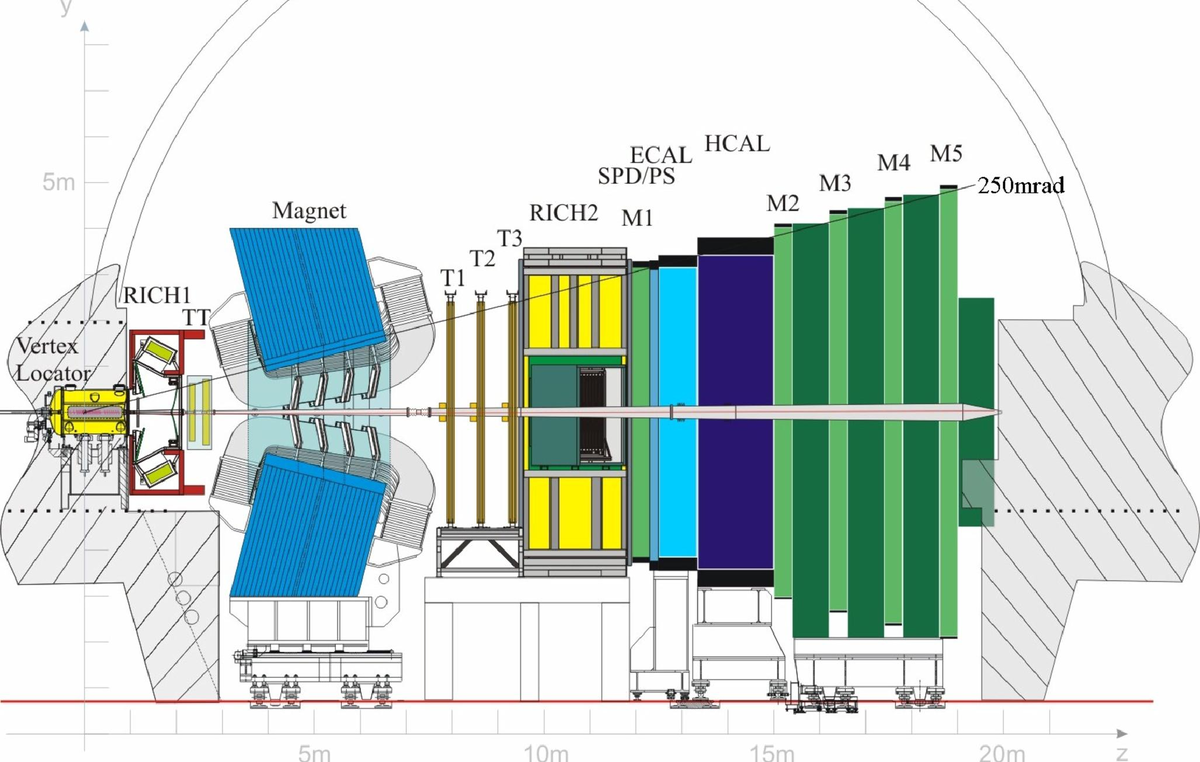}
  \caption{}
  \label{fig:LHCb_detector}
\end{subfigure}
\begin{subfigure}[b]{0.32\textwidth}
  \centering
  \includegraphics[width=\textwidth]{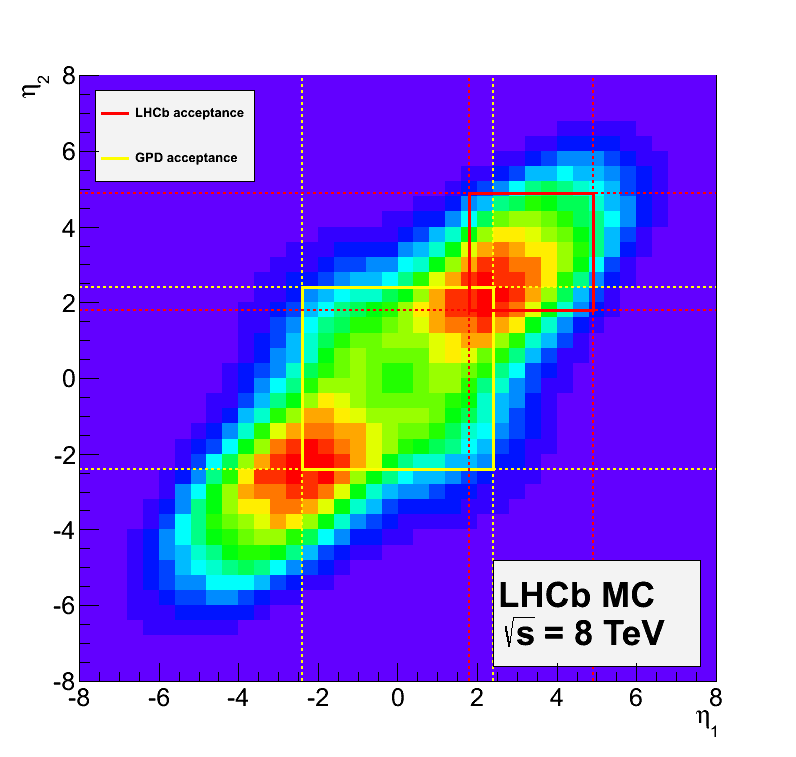}
  \caption{}
  \label{fig:LHCb_acc}
\end{subfigure}
\caption{Schematic view of the LHCb detector (a) and angular distribution of $b\bar b$ quark pairs produced from proton-proton collision at LHC (b)~\cite{LHCb_performances}. The red quadrant represents the LHCb angular acceptance.}
\end{figure}

The main purpose of the LHCb detector shown schematically in Figure~\ref{fig:LHCb_detector}, is the study of heavy meson decays. LHCb has good particle identification performances, from the two RICH detectors, the electromagnetic calorimeter and the muon stations, excellent momentum resolution ($\Delta p / p = 0.5\%$ at low momentum) and vertex identification (impact parameter resolution: $15+29/p_{T}\mathrm{[GeV]} ~\mathrm{\mu m}$) and good low momenta trigger performances, especially on di-muon tracks ($\sim 90\%$ of efficiency)~\cite{LHCb_performances}. LHCb is a forward detector, which covers a range of pseudorapidity $2<\eta<5$, with a large acceptance ($\sim 27\%$, Figure~\ref{fig:LHCb_acc}) for the wide variety of B mesons ($B^+$, $B^0$, $B_c$, $B_s$, $\Lambda_b$ etc.) produced from proton-proton collisions at LHC, whose decays constitute ideal probes for New Physics (NP).

\begin{figure}[htpb]
\centering
\begin{subfigure}[b]{0.37\textwidth}
  \centering
  \includegraphics[width=\textwidth]{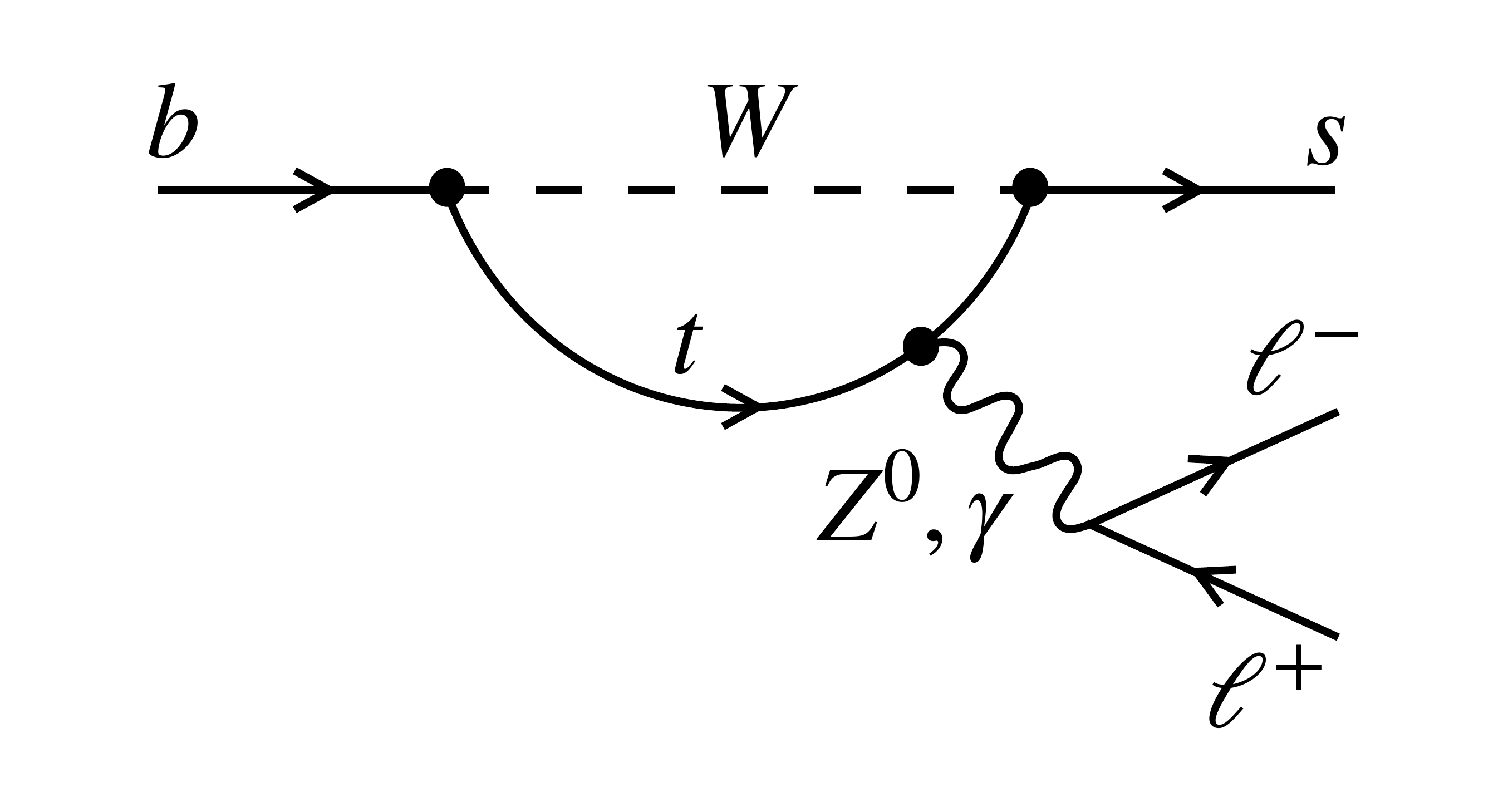}
  \caption{}
  \label{fig:bsll_fey}
\end{subfigure}
\begin{subfigure}[b]{0.29\textwidth}
  \centering
  \includegraphics[width=\textwidth]{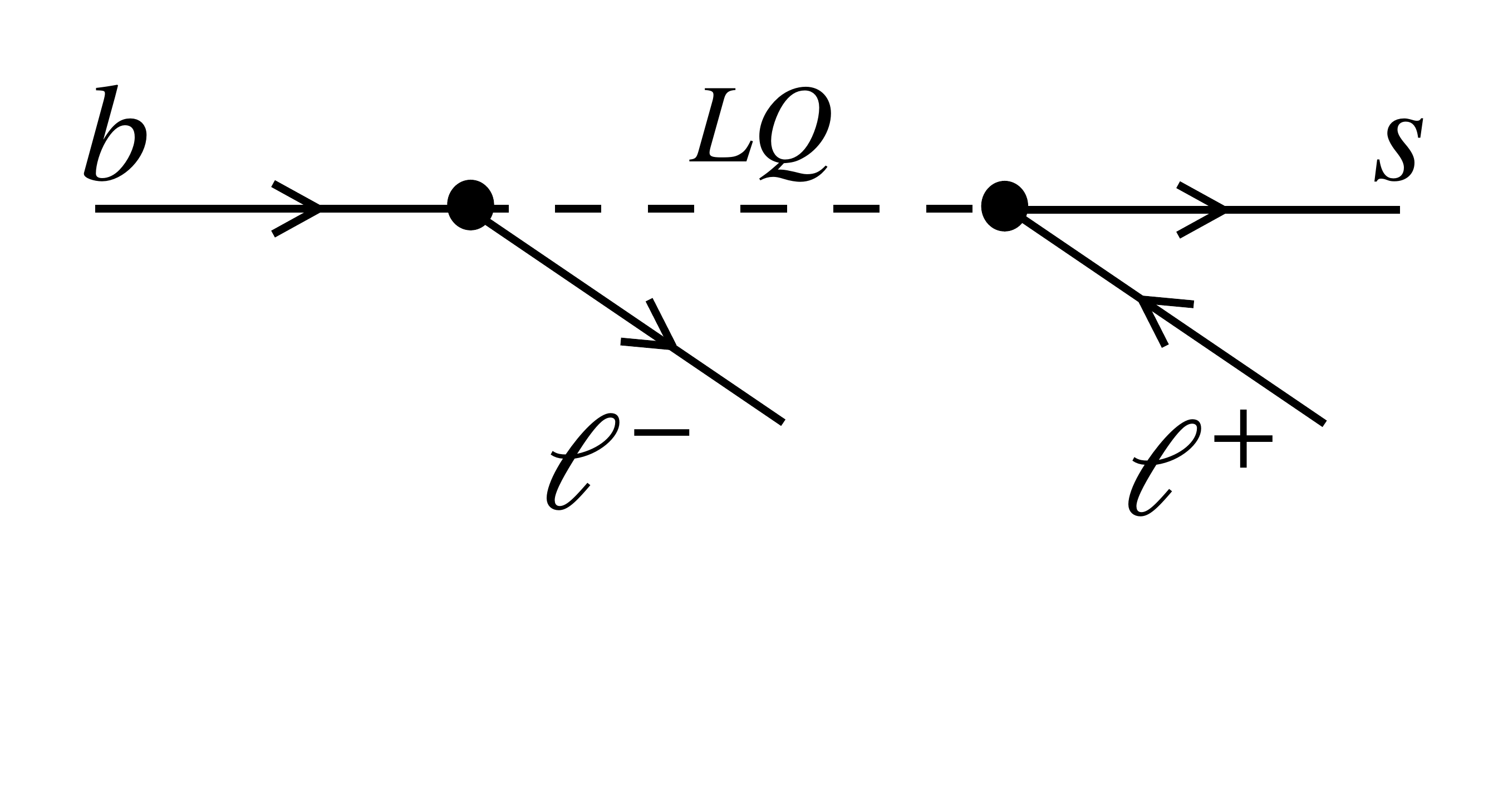}
  \caption{}
  \label{fig:LQ_fey}
\end{subfigure}
\caption{ (a) A SM contribution for $b\to s\ell\ell$ transition, happening via loop and involving the electroweak bosons $W$, $Z$ and $\gamma$. (b) A possible new physics contribution involving a leptoquark ($LQ$), which could couple directly quarks and leptons with different couplings for flavour families.}
\end{figure}

\noindent
The $B$ meson decays measurements are usually divided into two categories. One includes the processes which involve a transition from a $b$ quark to an $s$ quark, with the emission of a pair of leptons. These kinds of decays are flavour changing neutral currents and hence heavily suppressed by the SM: they are forbidden at tree-level and can only happen at higher orders, as shown in Figure~\ref{fig:bsll_fey}. Therefore, new physics mediators, leptoquarks~\cite{LQ_14,LQ_15,LQ_15_2,LQ_16} or $Z'$~\cite{Zprime_14,Zprime_15,Zprime_15_2,Zprime_15_3}, could modify significantly their amplitudes, adding new tree diagrams, as shown in Figure ~\ref{fig:LQ_fey}.

\noindent The second category includes $b\to c\ell\nu$ decays. These decays happen at tree level for the SM, and thus have larger branching ratios (up to few percent) than $b\to s\ell\ell$ decays ($\mathrm{BR} \sim 10^{-6}-10^{-7}$). Nevertheless, the presence of the neutrinos in the final states makes them experimentally challenging.

\subsection{\texorpdfstring{$\mathbf{b\to s\ell \ell}$}{fcnc} decays}
\label{sec:fcnc}
Several observables can be measured in $b\to s\ell \ell$ decays. Ratios of branching fractions of the form:

\begin{equation}
R_{K^*} = \frac{\mathcal{B}(B\to K^*\mu^+\mu^-)}{\mathcal{B}(B\to K^*e^+e^-)} \stackrel{\mathrm{SM}}{=} 1 \pm \mathcal{O}(10^{-2}),
\label{eq:RK}
\end{equation}

\noindent where the QCD uncertainties cancel out, are predicted by the SM with very high precision~\cite{RK_pred}, hence any deviation from the predicted value would be a clear sign of New Physics (NP). In addition, angular distributions of the final state particles could be also sensitive to NP contributions. Finally, NP interactions could also modify the decay rates depending on the lepton family considered.

\begin{figure}[htpb]
\centering
\includegraphics[width=0.35\textwidth]{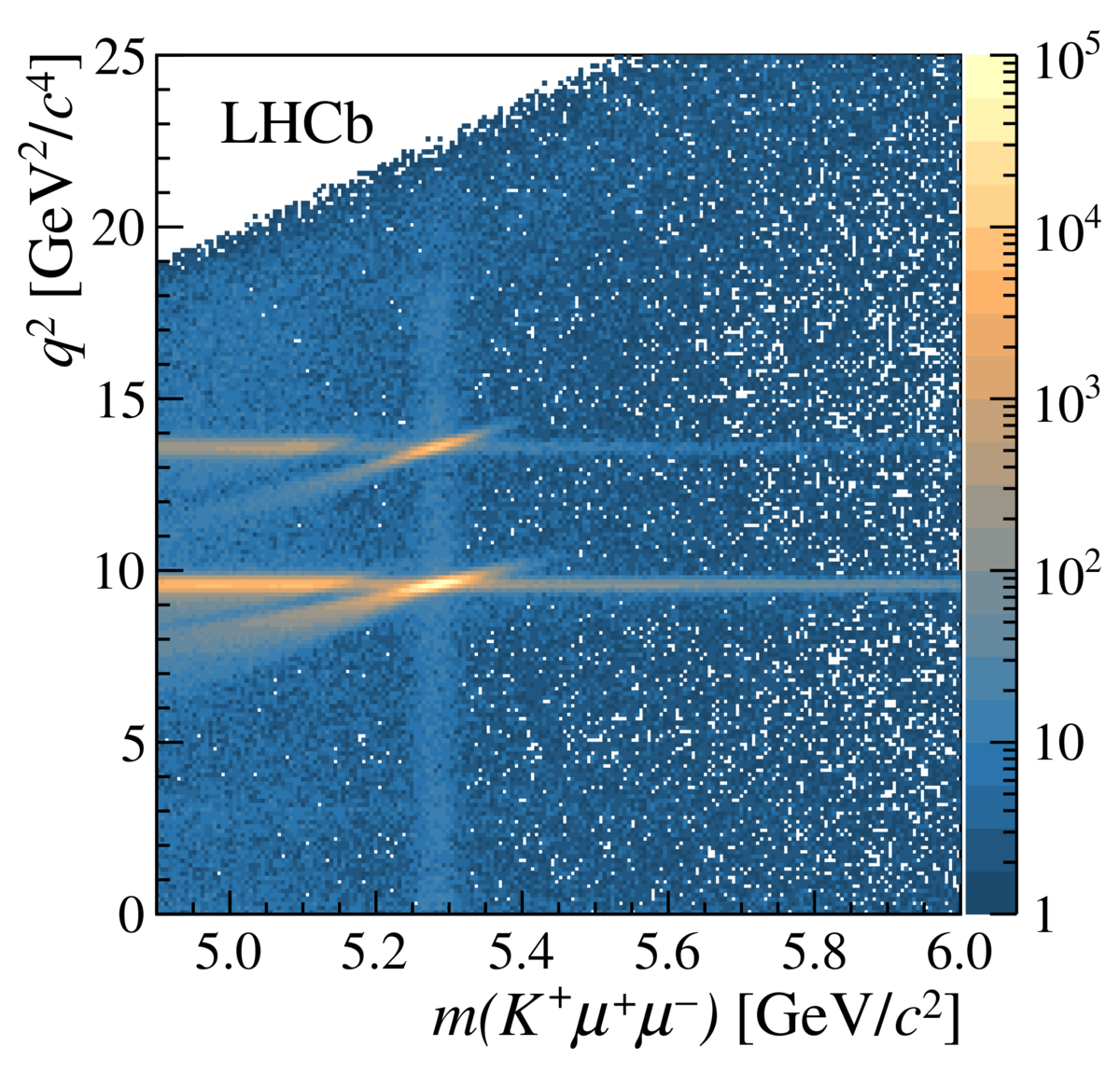}
\includegraphics[width=0.35\textwidth]{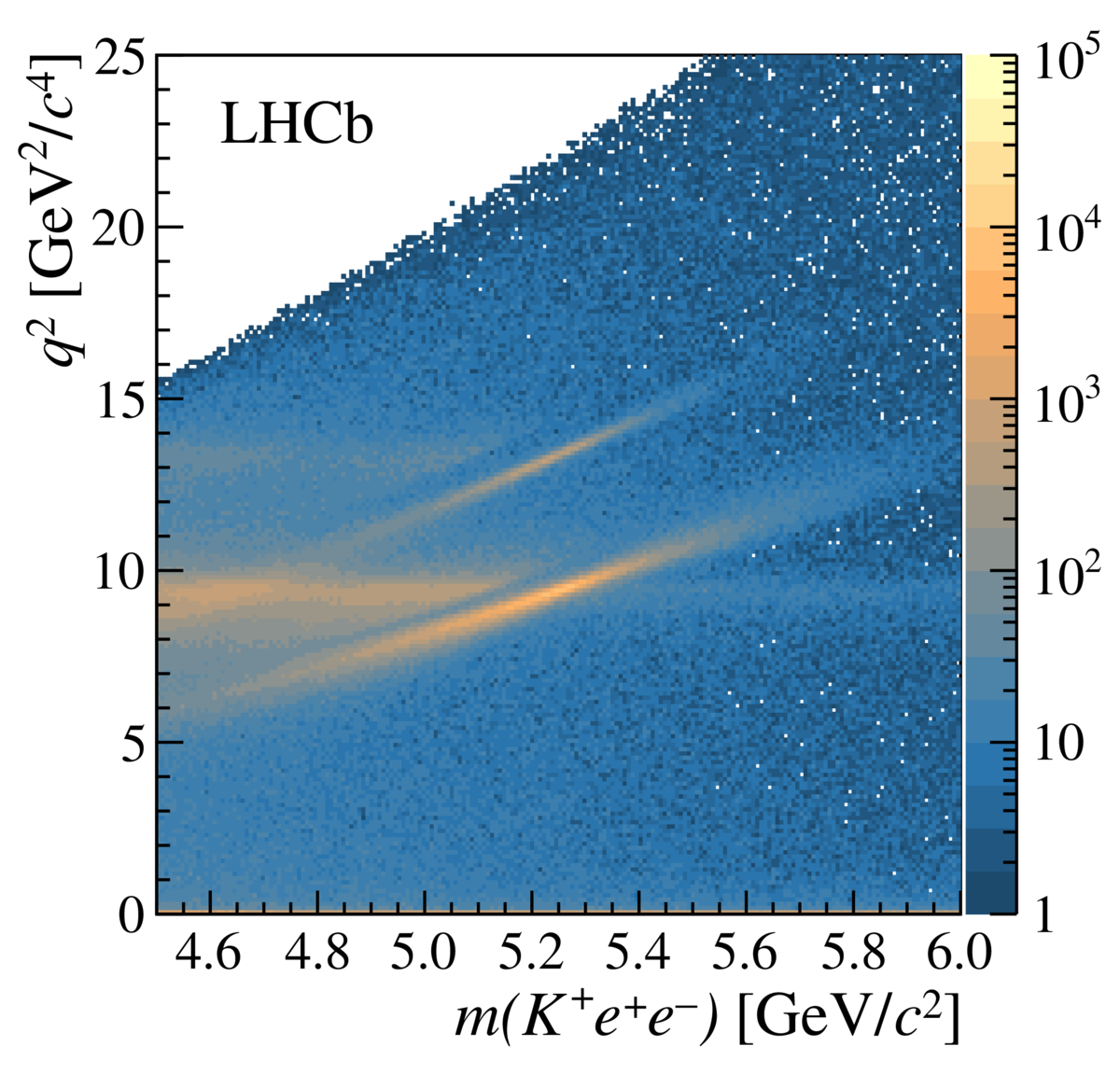}
\caption{Invariant mass of the final state particles for $B^+\to K^+ \mu^+\mu^-$ decays (left) and $B^+\to K^+ e^+e^-$ decays (right), for different values of the dilepton invariant mass squared ($q^2=m^2(\ell^+\ell^-)$). The decay $B^+\to K^+ \ell^+\ell^-$ should be visible as a vertical band in correspondence of the $B^+$ meson mass value, with two clear peaks pointing to the $J/\psi$ and $\psi(2S)$ resonance regions ($q^2 = m^2(J/\psi)$ and $q^2 = m^2(\psi(2S))$). From the right figure it is possible to understand how much the Bremsstrahlung effects impact the electron resolution in the whole $q^2$ range, where the vertical band around the $B^+$ mass vanishes and the tails of the resonances are much wider than for muons~\cite{RK_19}.}
\label{fig:resolution}
\end{figure}

\paragraph{LFU ratios}\mbox{} \\
\noindent
Relative rates as in Eq.~\ref{eq:RK} are usually experimentally measured as double ratios between the resonant (if the lepton pair comes from the decay of a $J/\psi$) and non resonant decay modes:

\begin{equation}
\begin{split}
R_{X} & = \frac{\mathcal{N}(B\to X \mu^+ \mu^-)}{\mathcal{N}(B\to X (J/\psi \to \mu^+ \mu^-))} \cdot \frac{\mathcal{N}(B\to X (J/\psi \to e^+e^-))} {\mathcal{N}(B\to X e^+e^-)} \\
 & \cdot \frac{\varepsilon(B\to X (J/\psi \to \mu^+ \mu^-))}{\varepsilon(B\to X \mu^+ \mu^-)} \cdot \frac{\varepsilon(B\to X e^+e^-)}{\varepsilon(B\to X (J/\psi \to e^+e^-))ß}
\end{split}
\label{eq:double_ratio}
\end{equation}

\noindent where $X$ can be a generic system containing an $s$ quark, $\mathcal{N}(B\to X \ell^+ \ell^-)$ and $\mathcal{N}(B\to X (J/\psi \allowbreak \to \ell^+\ell^-))$,  with $\ell = e,\mu$, are the yields of the rare and resonant decay modes obtained from fits to the invariant mass of the final state particles, and $\varepsilon(B\to X \ell^+ \ell^-)$ and $\varepsilon(B\to X (J/\psi \to \ell^+ \ell^-))$, with $\ell = e,\mu$, are the efficiencies of the four decays involved computed with MC simulation (usually calibrated with data from resonant channels). Since the $J/\psi$ satisfies lepton universality at order of tenths of percent~\cite{Jpsi_LFU}, the double ratio procedure is an efficient way to significantly reduce the systematic uncertainties arising from the differences in leptons reconstruction. In fact, while muons travel through the detector, most of the electrons emit bremsstrahlung photons before the magnet, making more challenging their energy estimation. In addition, electron reconstruction relies on information from the electronic calorimeter, which has higher occupancy and therefore higher energy thresholds. Electron reconstruction also suffers from worse momentum resolution as well as a lower tracking and PID efficiencies. All these differences affect the invariant mass distribution of the final state particles, both in statistics and in resolution, as it is shown in Figure~\ref{fig:resolution}.

\begin{figure}[htpb]
\centering
\includegraphics[width=0.35\textwidth]{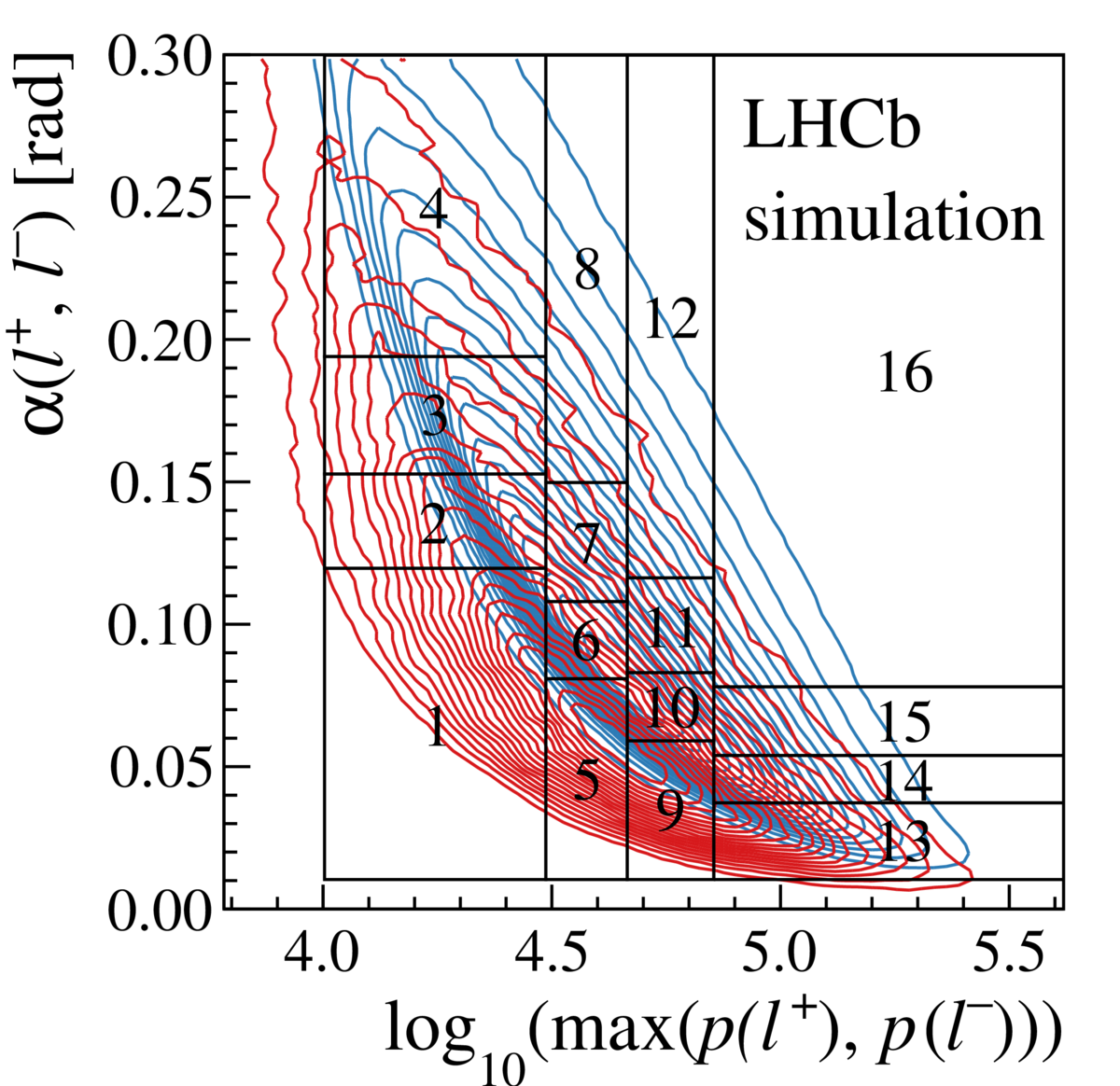}
\includegraphics[width=0.45\textwidth]{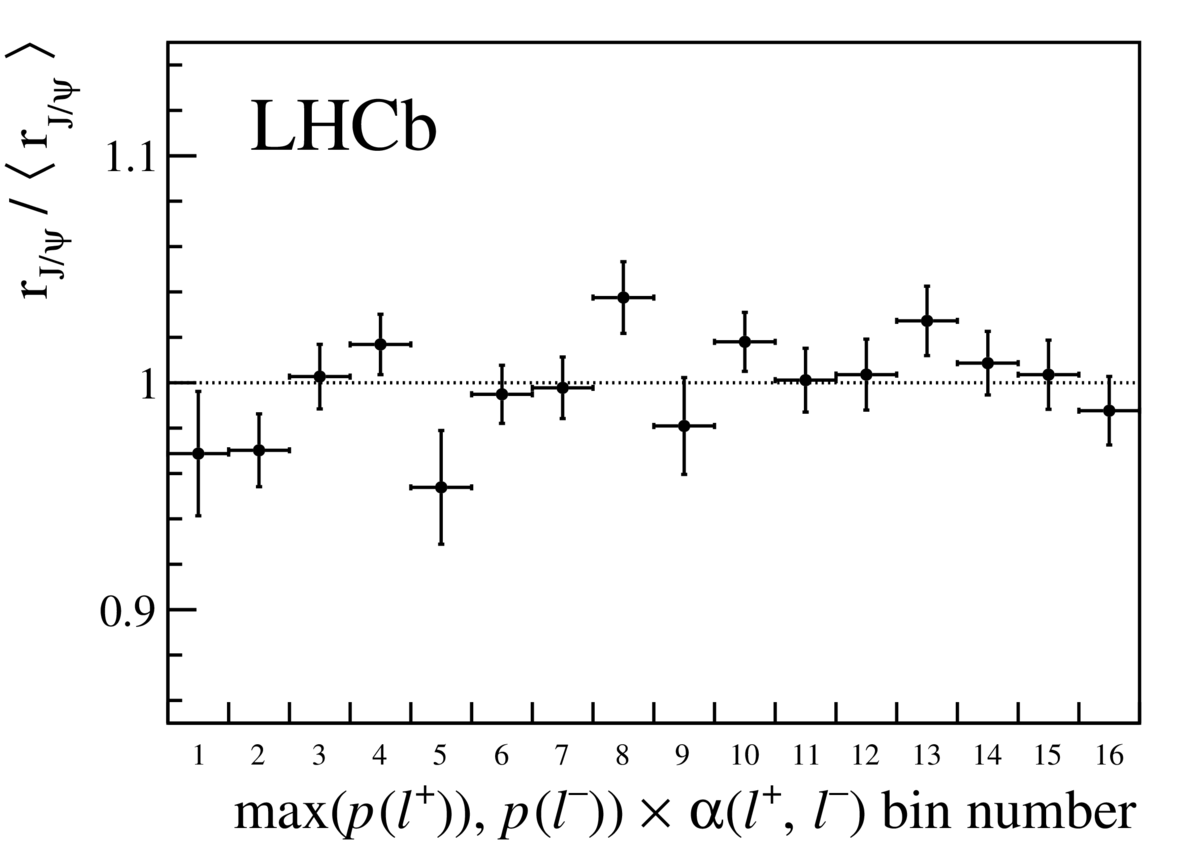}
\caption{Left: 2D distribution of the angle between the leptons and the maximum momentum of the leptons for $B^+ \to K^+ e^+e^-$ (red) and $B^+ \to K^+ (J/\psi \to e^+e^-)$ (blue) events. Right: $r_{J/\psi}$ value in bins of the same 2D distribution shown on the left. The $r_{J/\psi}$ value is checked to be in agreement with one and flat in all the bins.~\cite{RK_21}.}
\label{fig:rjpsi_flat}
\end{figure}

\noindent
In order to test that all these differences are under control, two powerful cross-checks are performed. The first one is the evaluation of the single ratio (and therefore more affected by $e/\mu$ differences):

\begin{equation}
r_{J/\psi} = \frac{\mathcal{B}(B\to X (J/\psi \to \mu^+\mu^-))}{\mathcal{B}(B\to X (J/\psi \to e^+e^-))}
\label{eq:rjpsi}
\end{equation}

\noindent
which is known to be equal to 1 with a precision of less than 1\%~\cite{Jpsi_LFU}. Furthermore, the absence of $r_{J/\psi}$ trends with respect to kinematical variables is also checked, as shown in Figure~\ref{fig:rjpsi_flat}. 

\noindent
The second cross-check makes use of both the $J/\psi$ and the $\psi(2S)$ resonances by constructing the double ratio:

\begin{equation}
R_{\psi(2S)} = \frac{\mathcal{B}(B\to X (\psi(2S) \to \mu^+\mu^-))}{\mathcal{B}(B\to X (J/\psi \to \mu^+\mu^-))} \cdot \frac{\mathcal{B}(B\to X (J/\psi \to e^+e^-))}{\mathcal{B}(B\to X (\psi(2S) \to e^+e^-))}
\label{eq:Rpsitwos}
\end{equation}

\noindent
which again is checked to be equal to one, and helps to verify the reduction of systematical uncertainties with the double ratio procedure.

\begin{figure}[htpb]
\centering
\includegraphics[width=0.40\textwidth]{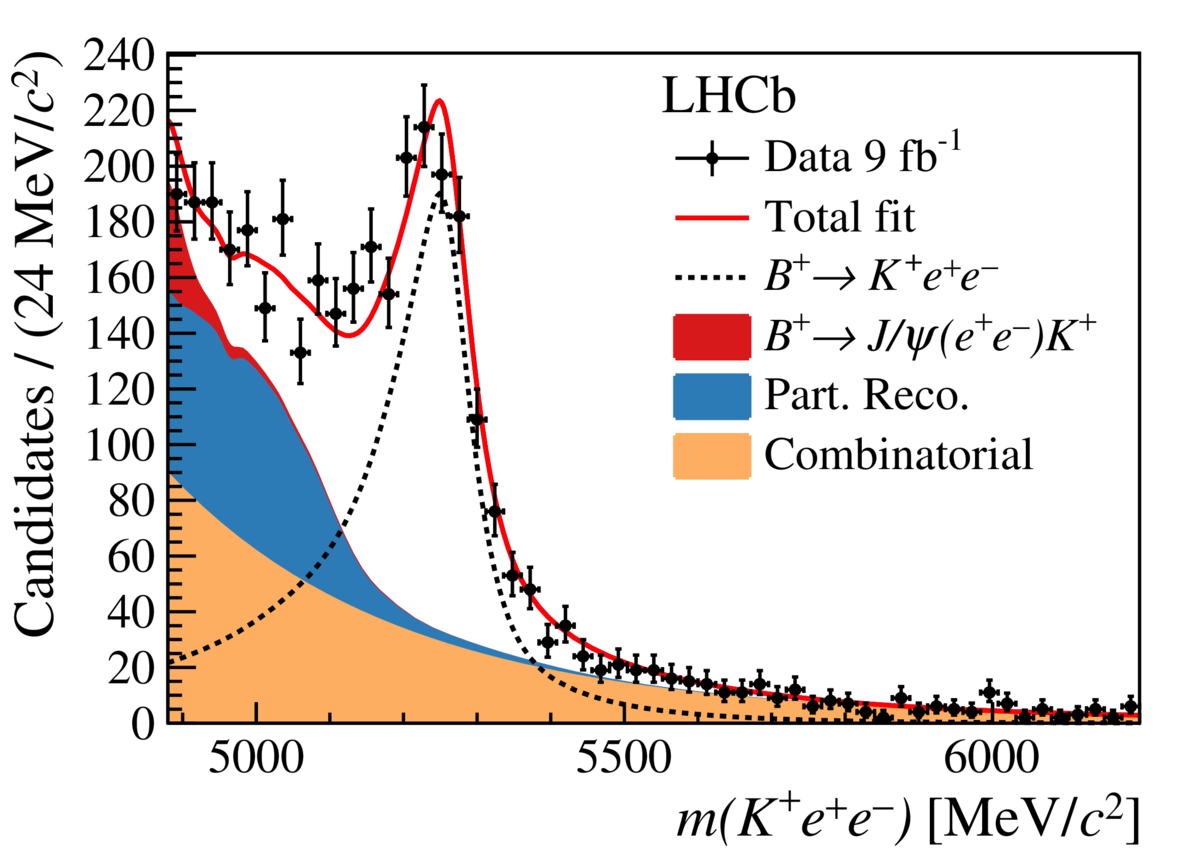}
\includegraphics[width=0.40\textwidth]{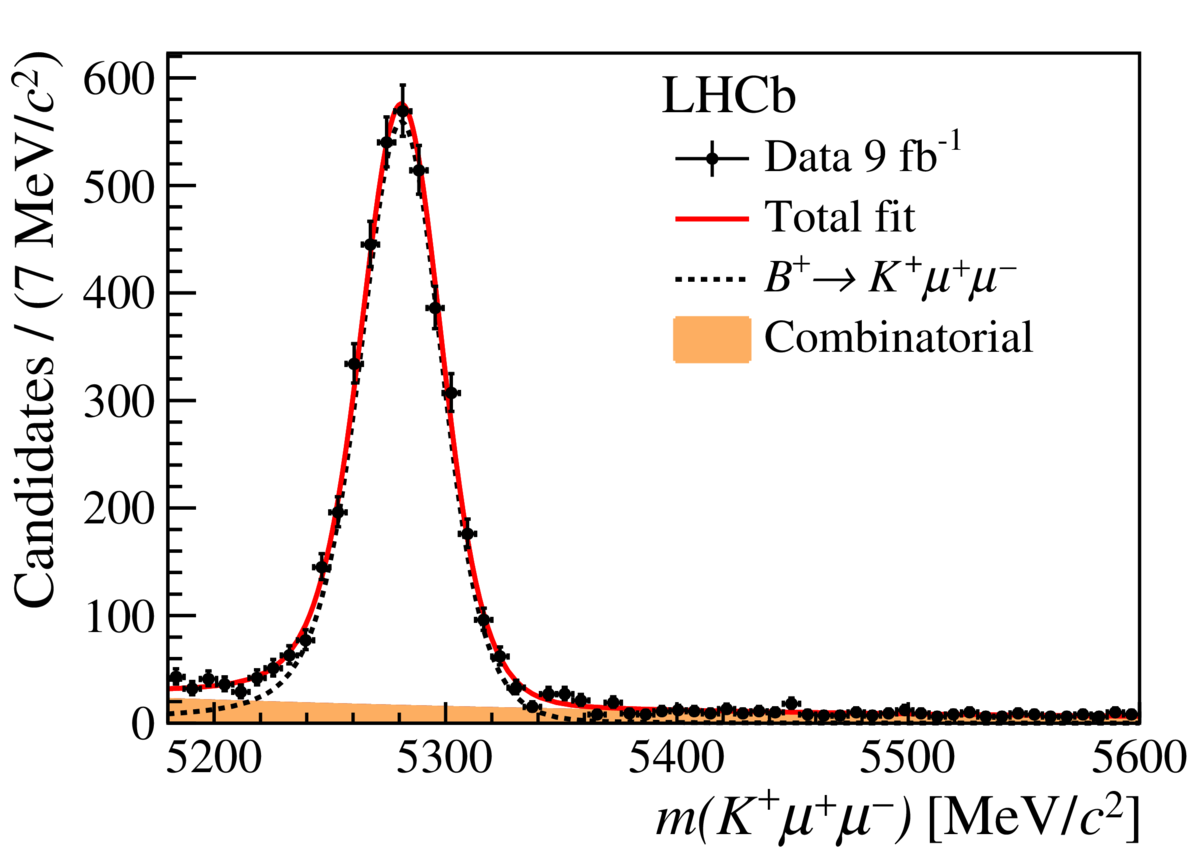}
\includegraphics[width=0.40\textwidth]{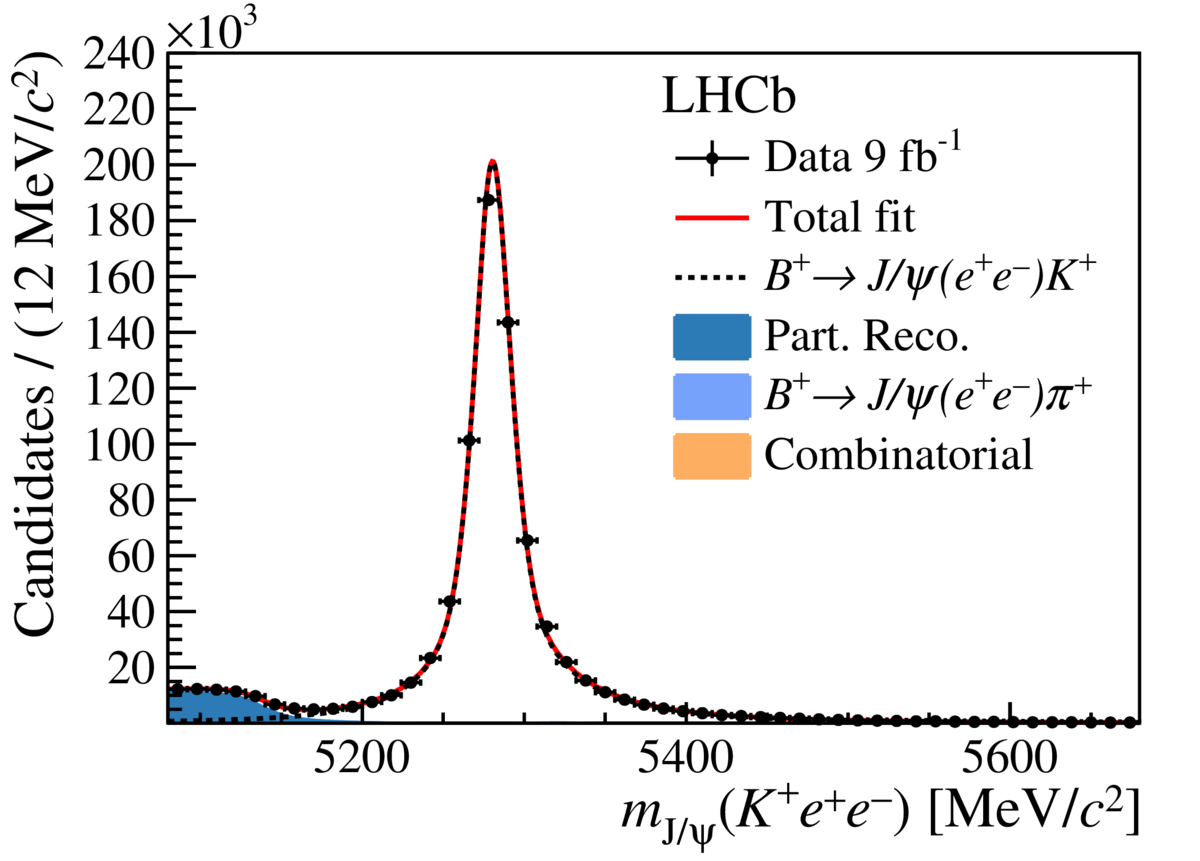}
\includegraphics[width=0.40\textwidth]{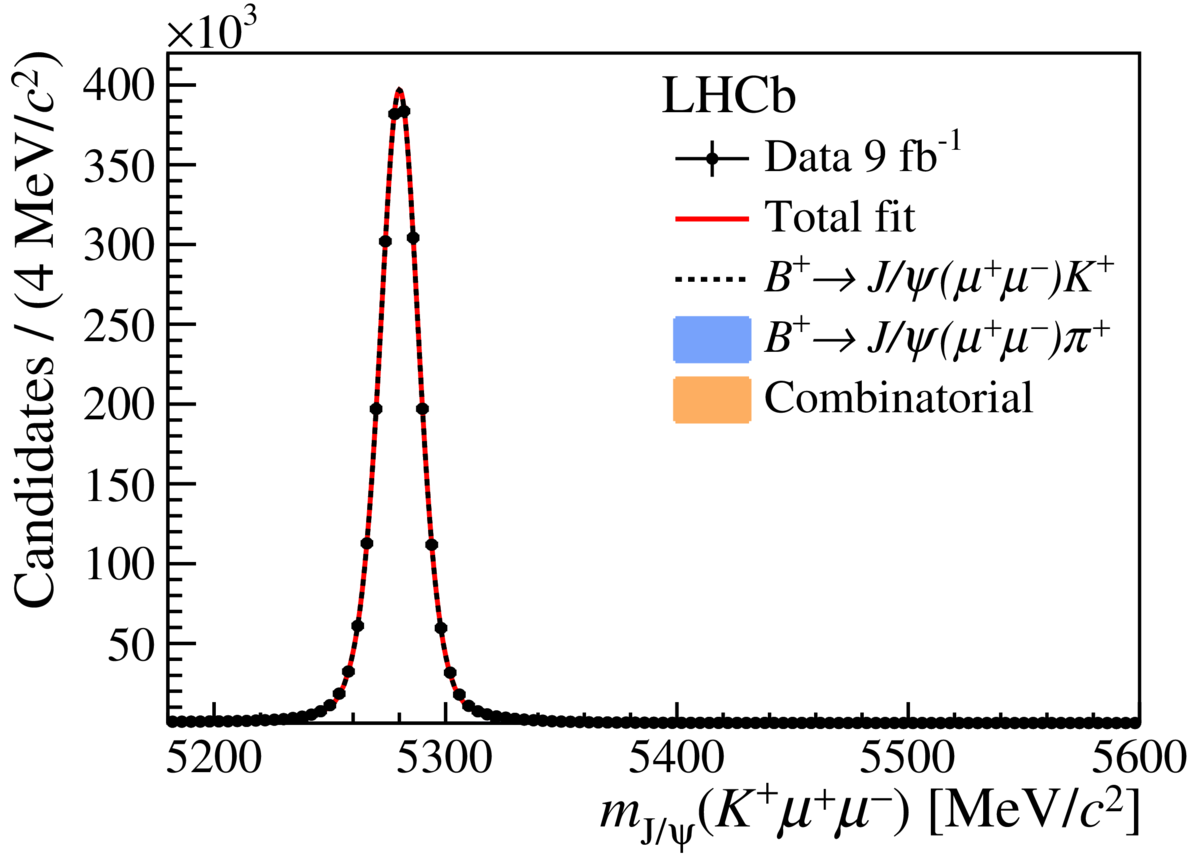}
\caption{Fit to the invariant mass of the final state particles for: $B^+ \to K^+e^+e^-$ (top left), $B^+ \to K^+\mu^+\mu^-$ (top right), $B^+ \to K^+(J/\psi \to e^+e^-)$ (bottom left), $B^+ \to K^+(J/\psi \to \mu^+\mu^-)$ (bottom right) decays~\cite{RK_21}. The measured yields from these fits are: $\mathcal{N}(B^+ \to K^+e^+e^-) = 1640\pm70$, $\mathcal{N}(B^+ \to K^+\mu^+\mu^-) = 3850\pm70$, $\mathcal{N}(B^+ \to K^+(J/\psi \to e^+e^-)) = 743300\pm900$ and $\mathcal{N}(B^+ \to K^+(J/\psi \to \mu^+\mu^-)) = 2288500\pm1500$.
}
\label{fig:RK_yields}
\end{figure}

LHCb has recently updated the measurement of $R_{K}$~\cite{RK_21} in a range of $1.1<q^2<6.\mathrm{GeV}^2/c^4$, analysing the complete set of data collected at the LHC (for a total of $9\mathrm{fb}^{-1}$ integrated luminosity). Figure~\ref{fig:RK_yields} shows the fits to the invariant mass of the final state particles for the four decay modes needed to evaluate the double ratio in Eq.~\ref{eq:double_ratio}.

\noindent
The $R_{K}$ value was found to be:

\begin{equation}
R_{K} (1.1<q^2<6 ~\mathrm{GeV}^2/c^4) = 0.846_{-0.039}^{+0.042}(\mathrm{stat})_{-0.012}^{+0.013}(\mathrm{sys})
\label{eq:RK_21_val}
\end{equation}

\noindent
which is compatible with the previous $R_K$ published value obtained with a smaller part of the data collected~\cite{RK_19}, and lays at 3.1 standard deviations from the SM predictions, representing the first evidence of LFU violation. The values of $r_{J/\psi}$ and $R_{\psi(2S)}$ were also found to be compatible with unity within one standard deviation:
\begin{align}
r_{J/\psi} = 0.981 \pm 0.020 (\mathrm{stat}+ \mathrm{sys}) \\
R_{\psi(2S)} = 0.997 \pm 0.011 (\mathrm{stat}+ \mathrm{sys})
\end{align}

\noindent
Deviations of LFU ratios from the SM predictions were already been observed by the LHCb collaboration looking at $B^0\to K^{*0}\ell^+\ell^-$ decays, with only $3\mathrm{fb}^{-1}$ of integrated luminosity. The measurement showed a $2.1 - 2.4\sigma$ tensions with the SM~\cite{RKst}, depending on the $q^2$ range considered:

\begin{equation}
  R_{K^{*0}} =
    \begin{cases}
      0.66^{+0.11}_{-0.07} \pm 0.03 & \mathrm{for}~(0.045<q^2<1.1 ~\mathrm{GeV}^2/c^4)\\
      0.69^{+0.11}_{-0.07} \pm 0.05 & \mathrm{for}~(1.1<q^2<6 ~\mathrm{GeV}^2/c^4)\\
    \end{cases}
\label{eq:RKst}    
\end{equation}

\noindent
Another LFU ratio measurement was also performed with $\Lambda^0_{b}\to pK^{-}\ell^+\ell^-$ decays, using $4.7\mathrm{fb}^{-1}$ of data. The result was found to be compatible with the SM, but with large uncertainties~\cite{RpK}.

\begin{equation}
R^{-1}_{pK}(0.1<q^2<6 ~\mathrm{GeV}^2/c^4) = 1.17 ^{+ 0.18}_{- 0.16} \pm 0.07 \\
\label{eq:RpK}
\end{equation}

\noindent
Since the workshop, the LHCb collaboration published an additional LFU ratio measurement, using $B^0\to K^0_S \ell^+ \ell^-$ and $B^+\to K^{*+} \ell^+ \ell^-$ decays~\cite{RKshort}, with the full dataset collected. The values of the ratios were found to be:

\begin{equation}
R_{K^0_S}(1.1<q^2<6 ~\mathrm{GeV}^2/c^4) = 0.66 ^{+ 0.20}_{- 0.14}(\mathrm{stat.}) ^{+ 0.02}_{- 0.04}(\mathrm{syst.}), \\
\label{eq:RKshort}
\end{equation}

\begin{equation}
R_{K^{*+}}(0.045<q^2<6 ~\mathrm{GeV}^2/c^4) = 0.70 ^{+ 0.18}_{- 0.13}(\mathrm{stat.}) ^{+ 0.03}_{- 0.04}(\mathrm{syst.}) ,\\
\label{eq:RKstplus}
\end{equation}

\noindent
in agreement with the SM predictions and with the previous tests of lepton universality.

\begin{figure}[htpb]
\centering
\includegraphics[width=0.44\textwidth]{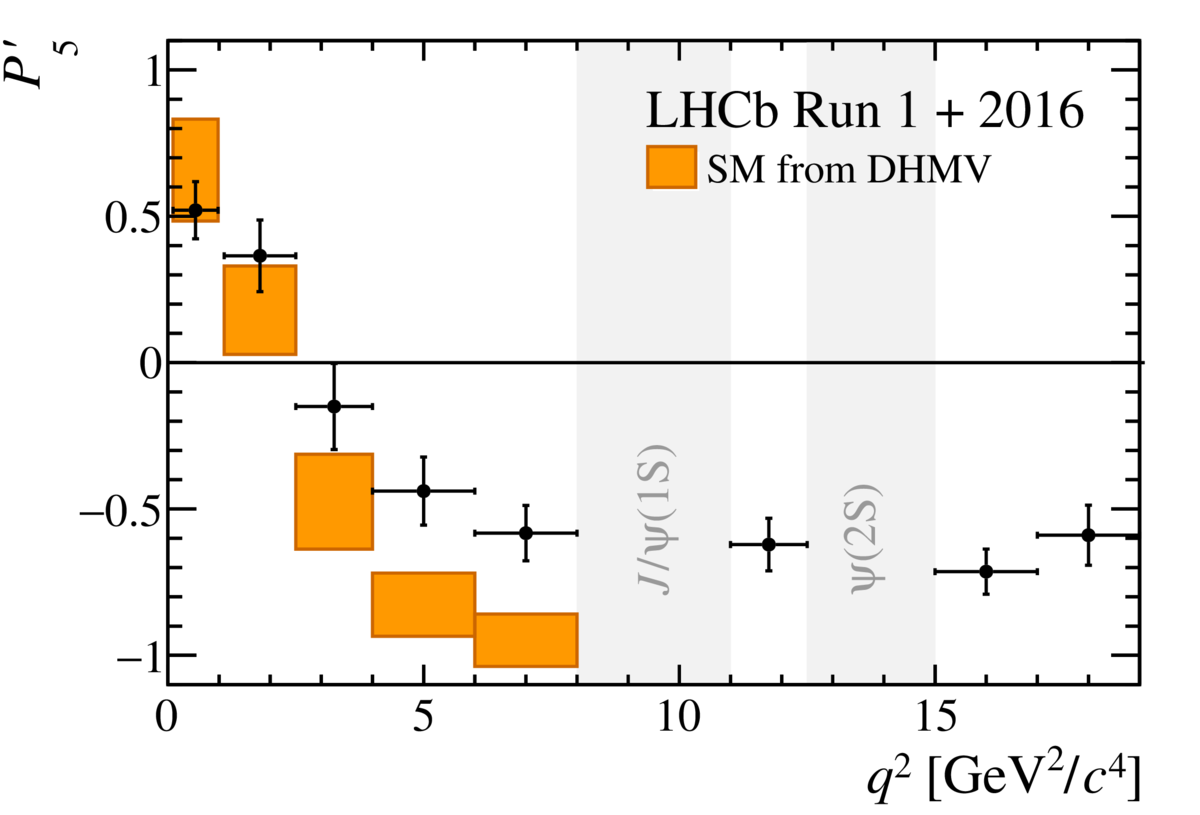}
\includegraphics[width=0.40\textwidth]{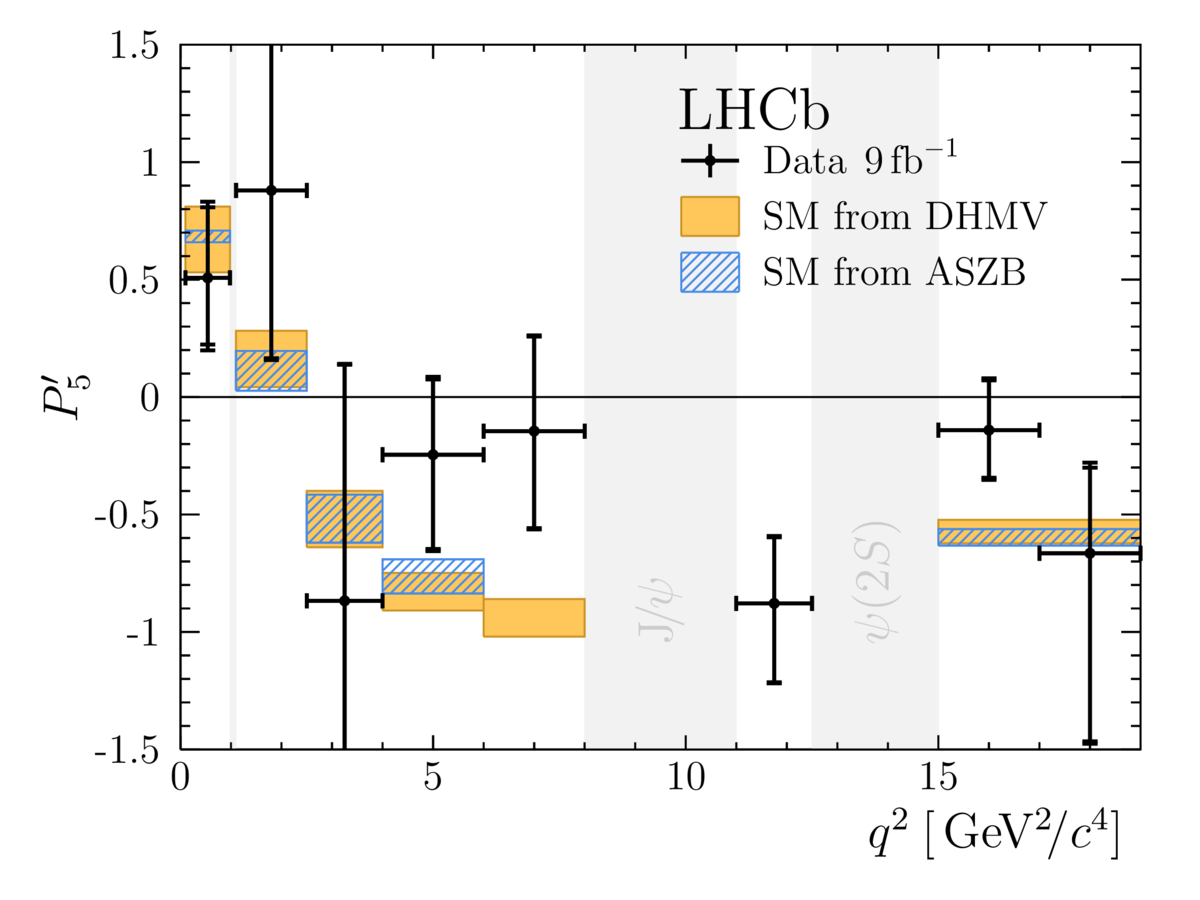}
\caption{$P'_5$ measured values for $B^0\to K^{*0}\mu^+\mu^-$~\cite{Angular_Kst0} (left) and for $B^+\to K^{*+}\mu^+\mu^-$~\cite{Angular_Kstplus} (right) decays in bins of the di-muon invariant mass. SM predictions are also shown.}
\label{fig:P5}
\end{figure}

\paragraph{Angular distributions}\mbox{} \\
The differential decay rate in bins of $q^2$ of processes such as $B^{0(+)}\to K^{*0(+)}\mu^+\mu^-$ can be fully described by $q^2$ and three angles: $cos(\theta_\ell)$, $cos(\theta_K)$ and $\phi$, as defined in~\cite{angular_rate}. The CP-averaged coefficients of the decay rate can be sensitive to New Physics contribution, but they have also challenging SM predictions, due to the modelling of hadronic form factors. The LHCb collaboration has measured the full set of CP-averaged angular observables in $B^0\to K^{*0}\mu^+\mu^-$ decays~\cite{Angular_Kst0}, with $6\mathrm{fb^{-1}}$ of integrated luminosity, and, more recently, in $B^+\to K^{*+}\mu^+\mu^-$ decays~\cite{Angular_Kstplus}, with the full dataset collected. Figure~\ref{fig:P5} shows the values measured by the two analysis for the $P'_{5}$ observable, specifically constructed in order to reduce QCD uncertainties at leading order, together with the respective SM predictions. The values measured with  $B^+\to K^{*+}\mu^+\mu^-$ decays were found to be in agreement with what observed in the previous $B^0\to K^{*0}\mu^+\mu^-$ analysis, confirming a local tensions with the SM predictions of 3.1$\sigma$, depending on the $q^2$ intervals considered and the hadronic uncertainties descriptions.

\noindent
Recently, the LHCb collaboration also updated the measurement of angular observables in $B^0\to \phi\mu^+\mu^-$ decays~\cite{Angular_phi}, using the full dataset collected. The observables were found to be overall in agreement with the SM predictions.

\begin{figure}[htpb]
\centering
\includegraphics[width=0.41\textwidth]{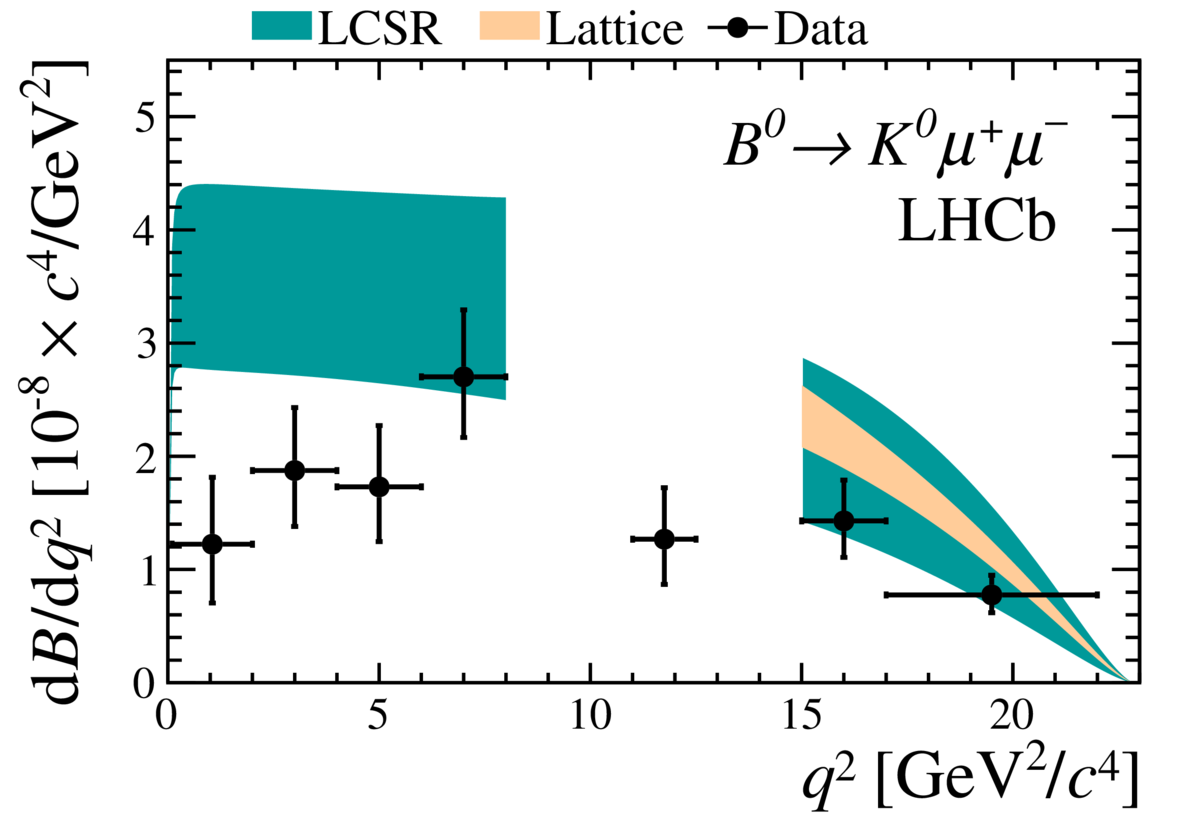}
\includegraphics[width=0.41\textwidth]{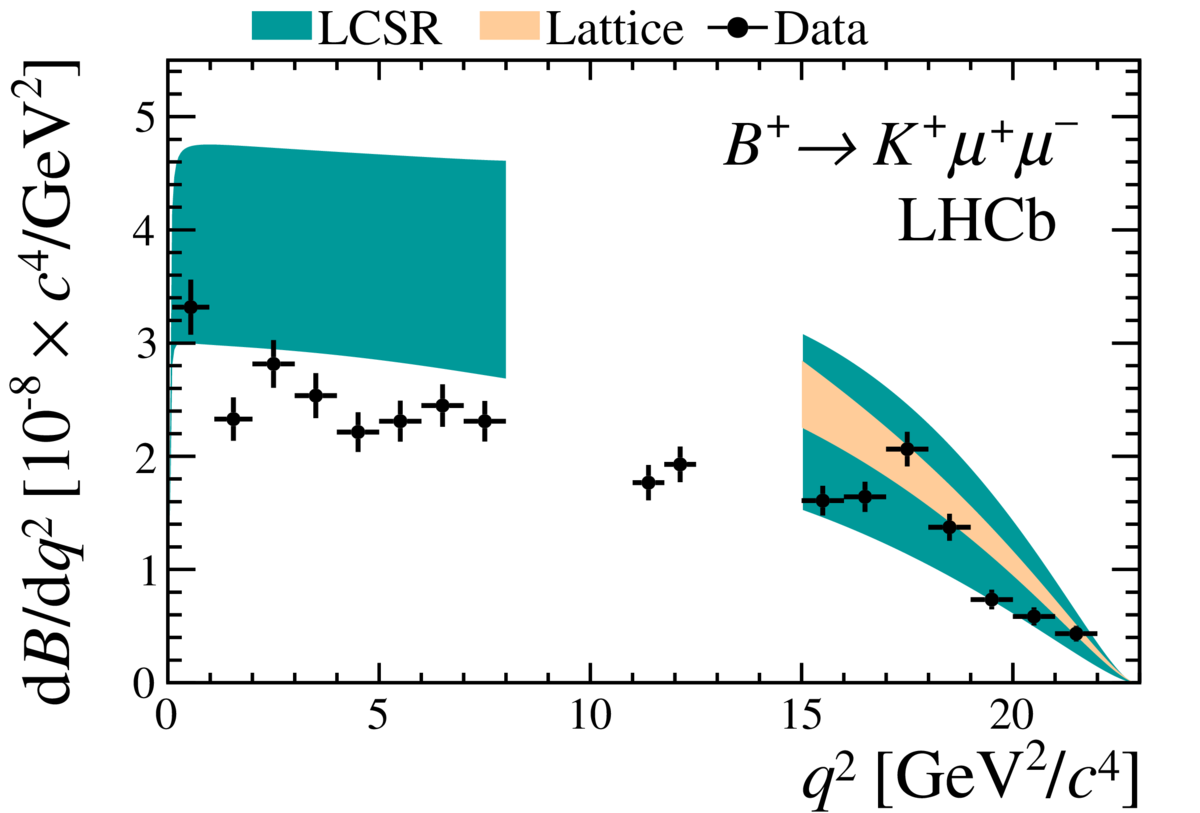}
\includegraphics[width=0.41\textwidth]{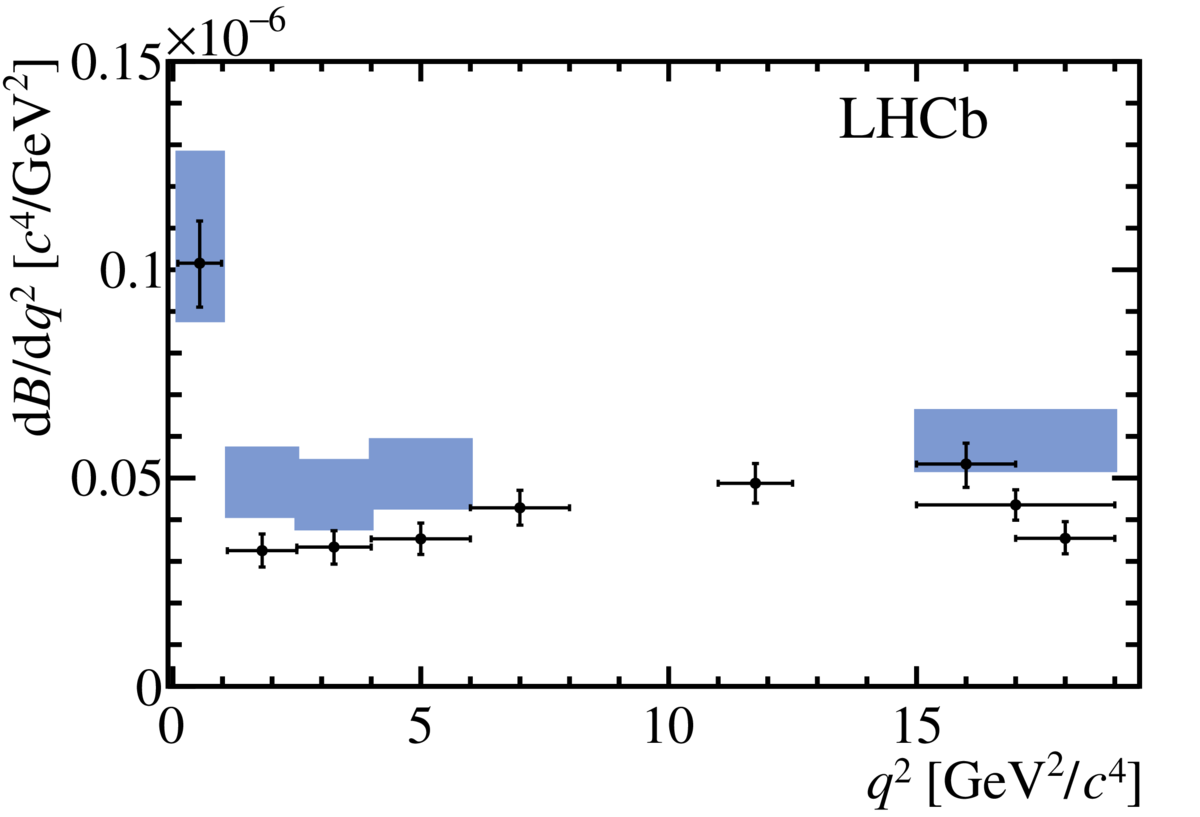}
\includegraphics[width=0.41\textwidth]{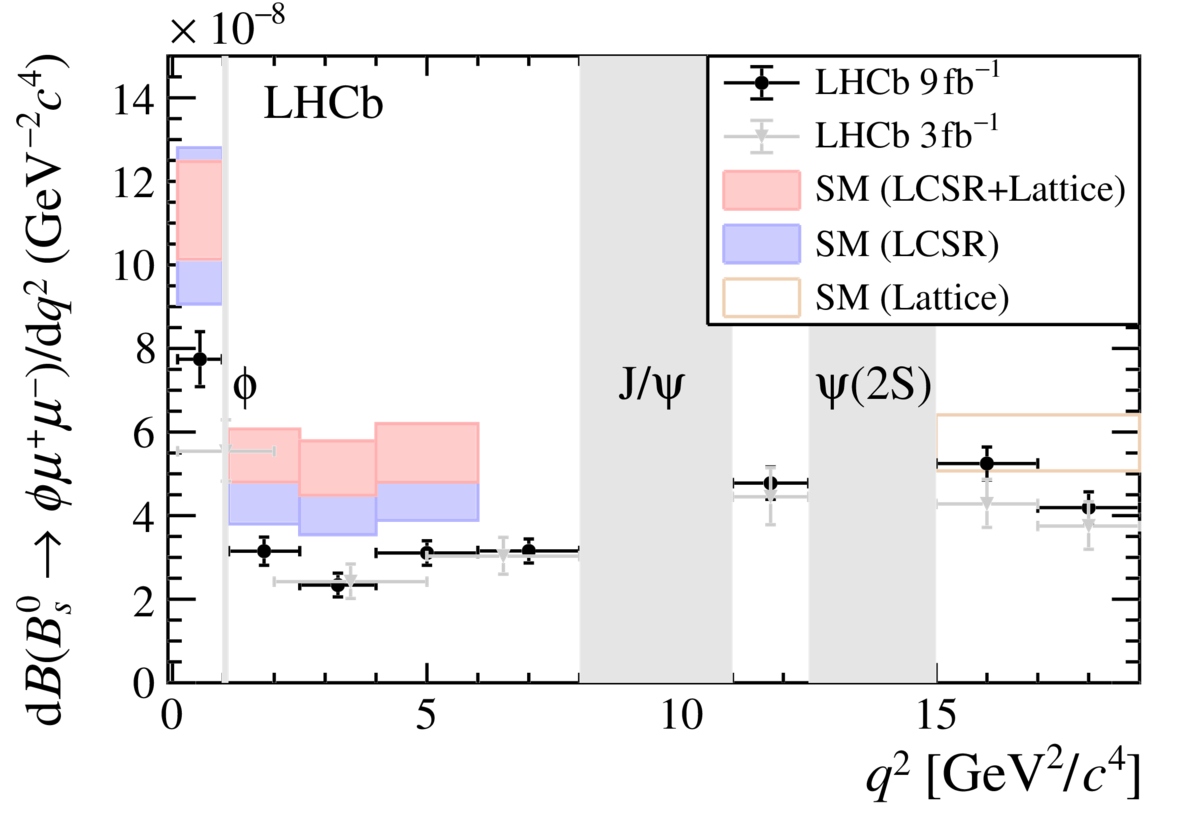}
\caption{Differential branching ratio measurements for $B^{0(+)}\to K^{0(+)}\mu^+\mu^-$ decays ~\cite{BR_Kst} (top), $B^{0}\to K^{*0}\mu^+\mu^-$~\cite{BR_Kplus} (bottom left), and $B^{0}_s\to \phi \mu^+\mu^-$~\cite{BR_phi} (bottom right) decays. SM predictions are also shown.}
\label{fig:BRs}
\end{figure}

\paragraph{\texorpdfstring{$\mathbf{b\to s \mu\mu}$}{bsmumu} branching ratios}\mbox{} \\
\noindent
Branching Ratio (BR) measurements are affected from large theory uncertainties, especially coming from the modelling of the hadronic form factors. The LHCb collaboration studied the differential branching fractions $dB/dq^2$ for several decay modes, shown in Figure~\ref{fig:BRs}. The measured branching fractions appear to be consistently lower than the SM predictions, with a larger discrepancy observed in $B_s^0\to\phi\mu^+\mu^-$decays~\cite{BR_phi}.

\paragraph{Interpretation}\mbox{} \\
\noindent
All the flavour anomalies described in the previous sections can be employed in so-called global fits, which allow to estimate the effective couplings (the {\textit{Wilson coefficients}) for the operators describing the $b\to s \ell\ell$ processes. Figure~\ref{fig:C9C10} shows a global fit to $C_9$ and $C_{10}$ coefficients, updated with the latest measurements of $R_K$~\cite{RK_21} and of the $\mathrm{BR}(B_s\to\mu\mu)$~\cite{Bmumu}, from the LHCb collaboration. All the measurements seem to point to a shift of these couplings with respect to the SM predictions. However, it has to be observed that the uncertainties of the $b\to s\mu\mu$ measurements are much larger than for the LFU ratios, mainly due to the description of hadronic form factors, and that there are ongoing discussion on whether this effect could be attributed to a wrong estimation of non-local charm-loop contribution, which could cause a similar shift of the $C_9$ SM predictions.

\begin{figure}[htpb]
\centering
\includegraphics[width=0.43\textwidth]{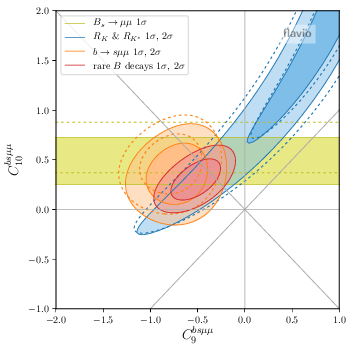}
\caption{Global fit to $C_9$ and $C_{10}$ Wilson coefficients including the latest LFU ratios (blue), the $b\to s\mu\mu$ measurements (orange) and the recent $\mathrm{BR}(B_s\to\mu\mu)$ (yellow). The combination is shown in red. The dashed lines indicate the constrains before the recent updates~\cite{C9_C10}.}
\label{fig:C9C10}
\end{figure}

\paragraph{\texorpdfstring{$b\to s\tau\tau$}{bstautau} status}\mbox{} \\
\noindent
Despite the fact that taus could be the most sensitive to new physic contributions, as shown in Figure~\ref{fig:BR_bstautau}, they are still largely unexplored due to the experimental challenges of having multiple neutrinos in the final states and, as is the case for the LHCb detector, of having a limited angular coverage. That is why experimental limits are still far away by several orders of magnitude from the SM predictions, see Figure~\ref{fig:BR_bstautau}.

\begin{figure}[htpb]
  \centering
  \includegraphics[width=0.4\textwidth]{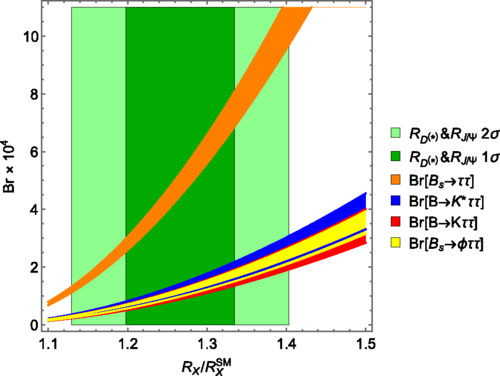}
  \qquad
  \begin{tabular}[c]{ccc}\hline
    Decay & SM predictions & Limits@90\% CL \\ \hline
    $B^0\to\tau\tau$ & $(2.22\pm0.19)\cdot10^{-8}$ & $<1.6\cdot10^{-4}$ (LHCb~\cite{BR_LHCb_Btautau}) \\
    $B_s^0\to\tau\tau$ & $(7.73\pm0.49)\cdot10^{-7}$ & $<5.2\cdot10^{-4}$ (LHCb~\cite{BR_LHCb_Btautau})\\
    $B^+\to K^+\tau\tau$ & $(1.20\pm0.12)\cdot10^{-7}$ & $<2.3\cdot10^{-3}$ (BaBar~\cite{BR_BaBar_BKtautau})\\
    \hline
  \end{tabular}
\caption{Top: Predictions of the branching ratios for $b\to s\tau\tau$ processes as a function of $R_{X}/R^{SM}_{X}$, where $X$ is a generic hadronic decay channel~\cite{BR_bstautau}. Bottom: experimental limits and relative SM predictions.}%
\label{fig:BR_bstautau}
\end{figure}

\subsection{\texorpdfstring{$\mathbf{b\to c\ell \nu}$}{cc} decays}
\label{sec:cc}
Similarly to what was already discussed for $b\to s\ell\ell$ decays, LFU ratios can be constructed as relative rates of $b \to c \tau \nu_\tau$ decays versus $b \to c \ell \nu_\ell$, where $\ell=e,\mu$. The missing energies from the neutrinos and the very busy environment of the LHC, where the initial $B$ meson momentum is unknown, make these measurements particularly challenging.

\paragraph{\texorpdfstring{$\mathbf{R(D^*)}$}{RDst}}\mbox{} \\
\noindent
The ratio $R_{D^*} = \mathcal{B}(B \to D^{*}\tau\nu) / \mathcal{B}(B \to D^{*}\mu\nu) (=  0.258 \pm 0.05$ for the SM), has been measured by LHCb using the $\tau$ {\textit{muonic}} decay $\tau^- \to \mu^- \bar{\nu_\mu}\nu_\tau$~\cite{RDst_muonic}, or the {\textit{hadronic}} one $\tau^- \to \pi^-\pi^+ \allowbreak \pi^-(\pi^0)\nu_{\tau}$~\cite{RDst_hadronic}. In the {\textit{muonic}} case, the signal and the normalization channels share the same visible final state, reducing significantly the experimental systematic uncertainties. In the {\textit{hadronic}} case, the $B\to D^{*-}\pi^+\pi^-\pi^+$ decay is instead used as a normalisation channel, in order to have the same visible final state as the signal one. This results in more uncertainty injected in the measurement, due to the limited knowledge of the $B\to D^{*-}\pi^+\pi^-\pi^+$ branching fraction. On the other hand, the use of the fully hadronic decay mode of the $\tau$ allows to precisely reconstruct the displaced $\tau$ vertex, the $B$ and $\tau$ line of flights and to have a better signal/background ratio, due to the absence of semileptonic background from multiple neutrinos. The values measured for $R(D^*)$ are:

\begin{equation}
R(D^*)_{{\textit{muonic}}} = 0.336 \pm 0.027(\mathrm{stat.}) \pm 0.030 (\mathrm{syst.})
\label{eq:RDst_mu}
\end{equation}

\begin{equation}
R(D^*)_{{\textit{hadronic}}} =0.291 \pm 0.019(\mathrm{stat.}) \pm 0.026(\mathrm{syst.}) \pm 0.013
\label{eq:RDst_had}
\end{equation}

\noindent
where the third uncertainty for the {\textit{hadronic}} measurement is due to the uncertainty on the

\noindent $\mathcal{B}(B\to D^{*-}\pi^+\pi^-\pi^+)$. The two measurements were found to be compatible, with the {\textit{muonic}} $R(D^*)$ value laying at $2.1\sigma$ above the SM predictions and the {\textit{hadronic}} one at $\sim1\sigma$ above the SM predictions, with larger uncertainties. Both measurements were performed using only $3\mathrm{fb}^{-1}$ of the data collected. The fits performed by the two analysis can be found in Figure~\ref{fig:RDst}.

\begin{figure}[htpb]
\centering
\includegraphics[width=0.80\textwidth]{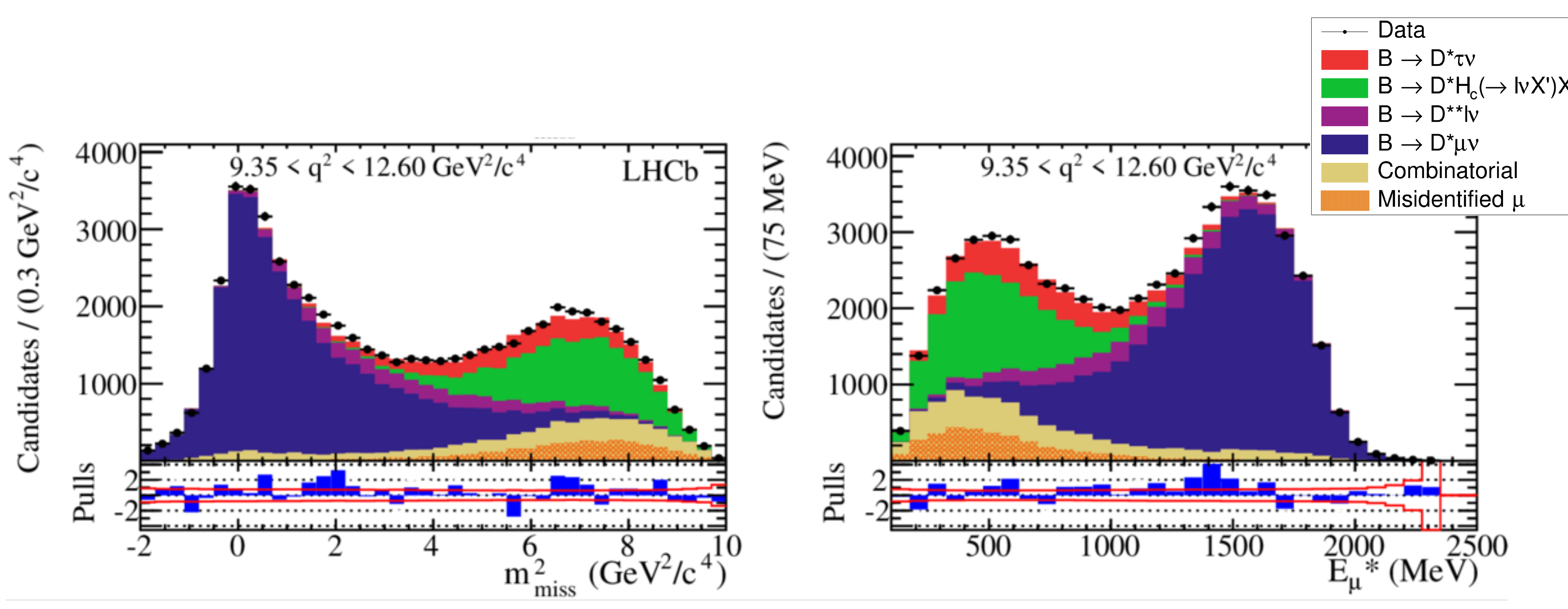}
\includegraphics[width=0.70\textwidth]{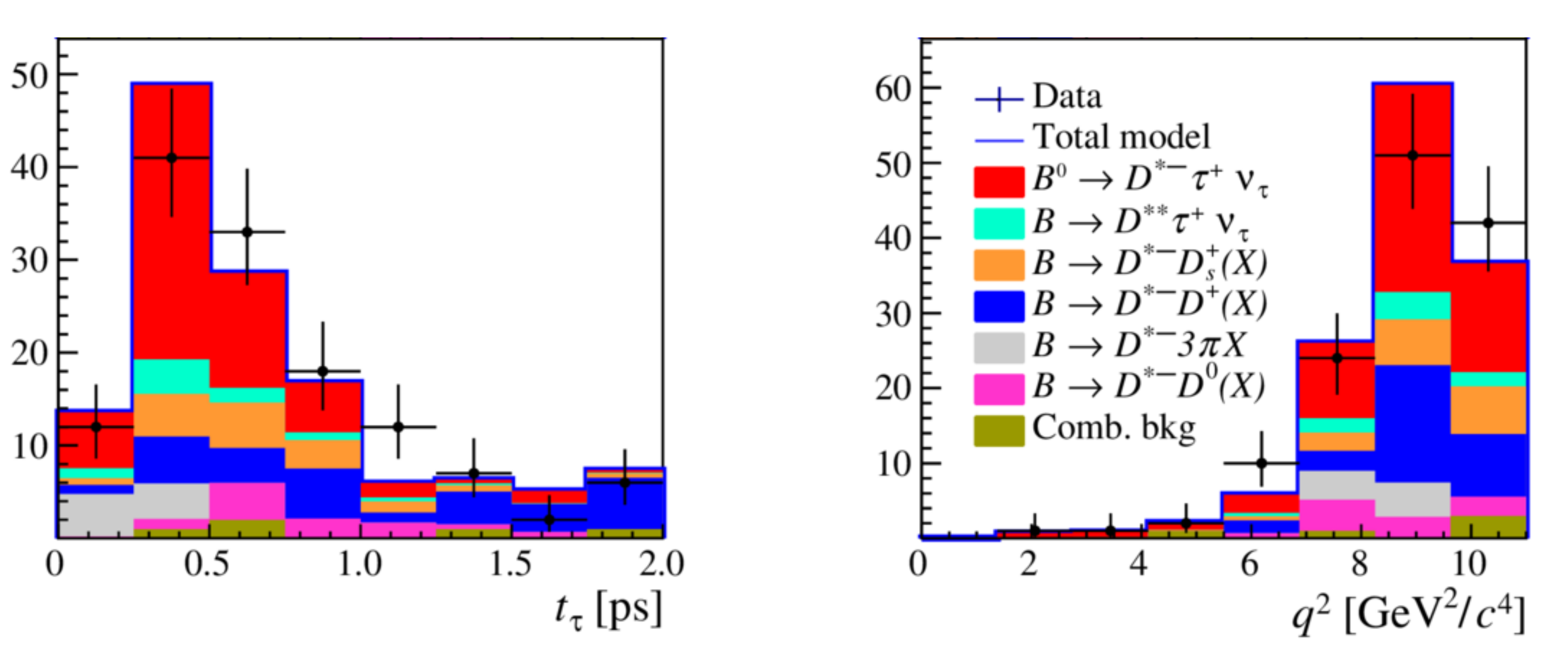}
\caption{Top: final fits for the {\textit{muonic}} $R_{D^*}$ measurement~\cite{RDst_muonic}. The fits shown are performed on the squared missing mass $m^2_{\mathrm{miss}} = (p_B - p_{D^*} -p_{\mu})^2$, on the energy of the muon in the $B$ meson mass frame, $E^*_\mu$, and in bins of $q^2 = (p_{B} - p_{D^*})^2$. Bottom: final fits for the {\textit{hadronic}} measurements of $R_{D^*}$~\cite{RDst_hadronic}. The fits shown are performed on the $q^2$ and on the $\tau$ decay time, in the last bin of the output of the multivariate analysis.}
\label{fig:RDst}
\end{figure}

\noindent
The combination of the $R(D^*)$ and $R(D)$ measurements, produced by the Heavy Flavour Averaging Group (HFLAV), is shown in Figure~\ref{fig:RDcomb}. The combination, which includes also measurements from the BaBar and Belle collaborations, exhibits a $3.1\sigma$ global tension with the SM predictions.

\begin{figure}[htpb]
\centering
\includegraphics[width=0.60\textwidth]{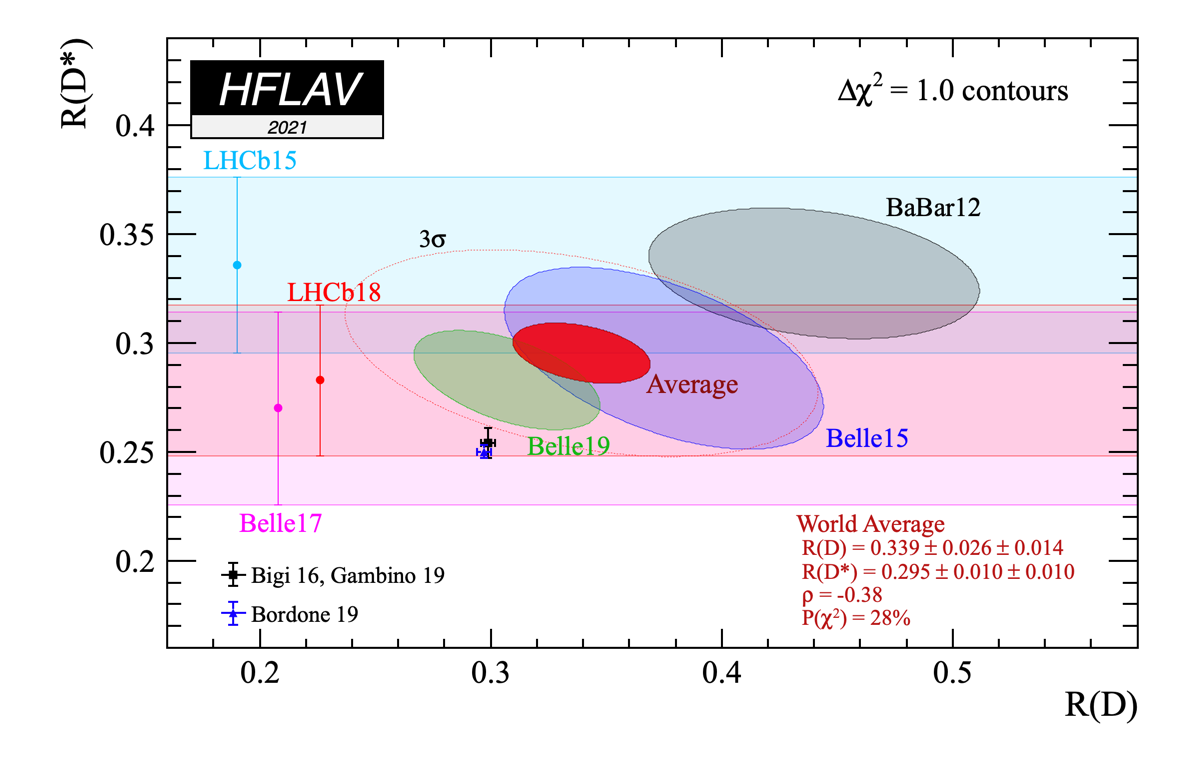}
\caption{$R(D^*)$ and $R(D)$ combination with inputs from LHCb, BaBar and Belle collaboration. The HFLAV average, in red, shows a $3.1\sigma$ tension with the SM prediction~\cite{HFLAV}.}
\label{fig:RDcomb}
\end{figure}

\paragraph{\texorpdfstring{$\mathbf{R(J/\psi)}$}{RJpsi}}\mbox{} \\
\noindent
LHCb performed an additional LFU ratios measurement, with a different spectator quark than $R_{D}$ and $R_{D^*}$, namely: $R_{J/\psi} = \mathcal{B}(B_c \to J/\psi\tau\nu) / \mathcal{B}(B_c \to J/\psi\mu\nu)$, for which the SM prediction lies in the interval [0.25,0.28], due to the choice of the model for form factors. In the analysis, which used only $3\mathrm{fb}^{-1}$ of the data collected, the $\tau$ lepton is reconstructed via its muonic decay and the final fit is performed on the squared missing mass, the $q^2$ and a combination of the energy of the muon in the $B$ meson mass frame and the decay time of the $B_c$. The value was found to be above the SM predictions of $2\sigma$:

\begin{equation}
R_{J/\psi} = 0.71\pm0.17(\mathrm{stat.})\pm0.18(\mathrm{syst.})
\label{eq:RJpsi}
\end{equation}

\noindent coherently with the $R_{D}$ and $R_{D^*}$ measurements.

\section{Lepton Flavour Violating decays}
\label{sec:LFV}

The LFU anomalies observed in $b\to s\ell\ell$ and in $b\to c\ell\nu$ decays described in this proceeding, renewed the interest in Lepton Flavor Violation (LFV) searches, since same new physics models (e.g. Leptoquark) could explain both LFU ratios and link them to a sizeable LFV~\cite{LFV_LQ}.

\paragraph{\texorpdfstring{$\mathbf{B^+\to K^+\mu^-\tau^{+}}$}{Kmutau}}\mbox{} \\
\noindent
LHCb recently updated the measurement of $B^+\to K^+\mu^-\tau^{+}$~\cite{Kmutau} using the full dataset collected and reconstructing the $B^+$ meson from the decay of a $B_{s2}^{*0} \to B^+(\to K^+\mu^-\tau) K^-$. In this way it was possible to reconstruct the $B^+$ flight direction and energy. A fit to the missing mass distribution in bins of the multivariate analysis output was performed, showing no visible excess. The limit set on this decay was: $\mathcal{B}(B^+ \to K^+ \mu^- \tau^+ ) < 3.9(4.5) \times 10^{-5}$ at 90\%(95\%) of confidence level, comparable with the world best limit.

\paragraph{\texorpdfstring{$\mathbf{B^+\to K^+\mu^{\pm}e^{\mp}}$}{Kmue}}\mbox{} \\
\noindent
In this search, the tree-level decays $B^+\to K^+ J/\psi(\to\ell^+\ell^-)$, with $\ell=(e,\mu)$ were used as normalization control channels. The LHCb analysis for $B^+\to K^+\mu^{\pm}e^{\mp}$ used only $3\mathrm{fb}^{-1}$ of the data collected and showed no significant signal, but improved the current best limits on these decays by more than one order of magnitude~\cite{Kmue}, finding: $\mathcal{B}(B^+ \to K^+ \mu^- e^+ ) < 7.0(9.5) \times 10^{-9}$ $\mathcal{B}(B^+ \to K^+ \mu^+ e^- ) < 6.4(8.8) \times 10^{-9}$ both at 90\%(95\%) of confidence level.

\paragraph{\texorpdfstring{$\mathbf{B^0_s\to \tau^{\pm}\mu^{\mp}}$}{mutau}}\mbox{} \\
\noindent
Searches for $B^0_s\to \tau^{\pm}\mu^{\mp}$ decays by the LHCb collaboration used the fully hadronic decay of the $\tau$ ($\tau^- \to \pi^-\pi^+\pi^-(\pi^0)\nu_{\tau}$). With only $3\mathrm{fb}^{-1}$ of integrated luminosity, no excess was found in the reconstructed $B$ meson mass distribution\cite{mutau}. Nevertheless the two world-best limits for these decays were set: $\mathcal{B}(B^0 \to \tau^{\pm}\mu^{\mp}) < 1.2(1.4) \times 10^{-5}$ and  $\mathcal{B}(B^0_s \to \tau^{\pm}\mu^{\mp}) < 3.4(4.2) \times 10^{-5}$, both at 90\%(95\%) of confidence level.

\paragraph{\texorpdfstring{$\mathbf{B^0_s\to \mu^{\pm}e^{\mp}}$}{mue}}\mbox{} \\
\noindent
The strongest limits on $B^0_s\to \mu^{\pm}e^{\mp}$ were set by the LHCb collaboration in~\cite{mue}, using only $3\mathrm{fb}^{-1}$ of integrated luminosity and looking at the lepton-pair invariant mass distribution in bins of the Boosted Decision Tree output. The limits were found to be $\mathcal{B}(B^0_s \to e^{\pm}\mu^{\mp}) < 6.3(5.4)\times 10^{-9}$ and $\mathcal{B}(B^0_s \to e^{\pm}\mu^{\mp}) < 7.2(6.0)\times 10^{-9}$ at 90\%(95\%) of CL, for the hypotheses of an amplitude completely dominated by the heavy eigenstate and by the light eigenstate respectively. Best-world upper limits were also found for $\mathcal{B}(B^0 \to e^{\pm}\mu^{\mp}) < 1.3(1.0)\times 10^{-9}$ at 90\%(95\%) of confidence level.

\paragraph{\texorpdfstring{$\mathbf{D_{(s)}^+\to h^{\pm}\ell^+\ell^{'\mp}}$}{Dhll}}\mbox{} \\
\noindent
The LHCb collaboration has recently performed a search for 25 rare and forbidden charm decays of the form $D_{(s)}^+\to h^{\pm}\ell^+\ell^{'\mp},$ with $\ell=e,\mu$ and $h=\pi,K$, using $1.6\mathrm{fb}^{-1}$ of integrated luminosity. No signal evidence was observed for the 25 decays modes and limits on the branching fractions were set between $1.4 \times 10^{-8}$ and $6.4 \times 10^{-6}$ at 90\% of confidence level~\cite{Dhll}.

\section{Conclusion}
\label{sec:conclusion}
Flavour physics is currently one of the most interesting area for New Physics searches. Numerous flavour anomalies were observed by the LHCb collaboration in ratios of branching fractions, angular observables and muon decay rates. All the measurements seem to point to a coherent pattern, with a shift of the effective coupling ({\textit{Wilson coefficient}}) from the SM predictions, that could be explained by several New Physics models (Leptoquark or a new heavy gauge boson such as $Z'$).

\noindent
Flavour anomalies could be confirmed or disproved in a very near future. Several LHCb measurements are in preparation: the updates of the aforementioned measurements with the full data collected by LHCb from 2011 to 2018; LFU ratios with different hadronic final states, such as $R_{K\pi\pi}$, $R_{\phi}$ for $b\to s\ell\ell$, or $R_{D_s}$, $R_{\Lambda_c}$ for $b\to c\ell\nu$; simultaneous measurements of different hadronic final states, such as $R_K \& R_K^*$ or $R_D \& R_D^*$; and the analysis of the electron angular distributions in $B^{0(+)}\to K^{*0(+)}e^+e^-$ decays.

\noindent
In addition, with the starting of the Run3 data taking period (2022-2025) at LHC, the data collected are expected to become approximately three times larger in three years. This increase in statistics will help both to reduce the statistical uncertainties (the main limitation in most of the analysis presented) and the systematics due to data-driven models, leading to an unprecedented precision for flavour measurements. Furthermore, the LHCb detector will undergo several staged upgrades in the following years, where the removal of the hardware trigger and the replacements of several sub detectors, such as the vertex and the tracking detectors, will reduce the background coming from charged and neutral tracks and will make the electronic (and even tauonic) modes more accessible. 
Belle II work will also be of fundamental importance in order to independently clarify the flavour anomalies that have been puzzling the physics community in the last decade.

\bibliography{mybib.bib}

\begin{thebibliography}{10}
\providecommand{\url}[1]{\texttt{#1}}
\providecommand{\urlprefix}{URL }
\expandafter\ifx\csname urlstyle\endcsname\relax
  \providecommand{\doi}[1]{doi:\discretionary{}{}{}#1}\else
  \providecommand{\doi}{doi:\discretionary{}{}{}\begingroup
  \urlstyle{rm}\Url}\fi
\providecommand{\eprint}[2][]{\url{#2}}

\bibitem{Jpsi_LFU}
{Ablikim M. et al, [BESIII Collaboration]},
\newblock \emph{Precision measurements of {$B[\psi(3686)\to\pi^+\pi^-J/\psi]$
  and $B[J/\psi\to\ell^+\ell^-]$}},
\newblock Physical Review D \textbf{88}(3) (2013),
\newblock \doi{10.1103/physrevd.88.032007}.

\bibitem{K_LFU}
{Lazzeroni C. et al, [NA62 collaboration]},
\newblock \emph{Precision measurement of the ratio of the charged kaon leptonic
  decay rates},
\newblock Physics Letters B \textbf{719}, 326 (2013),
\newblock \doi{10.1016/j.physletb.2013.01.037}.

\bibitem{pi_LFU}
{Aguilar-Arevalo et al, [PiENu Collaboration]},
\newblock \emph{{Improved Measurement of the
  $\ensuremath{\pi}\ensuremath{\rightarrow}\mathrm{e}\ensuremath{\nu}$
  Branching Ratio}},
\newblock Phys. Rev. Lett. \textbf{115}, 071801 (2015),
\newblock \doi{10.1103/PhysRevLett.115.071801}.

\bibitem{PDG}
{Zyla, P.A. et al [PArticle Data Group]},
\newblock \emph{Review of particle physics},
\newblock PTEP \textbf{2020}(8), 083C01 (2020),
\newblock \doi{10.1093/ptep/ptaa104}.

\bibitem{Z_LFU}
{S. Schael et al. [ALEPH and DELPHI and L3 and OPAL and SLD Collaborations and
  LEP Electroweak Working Group and SLD Electroweak Group and SLD Heavy Flavour
  Group]},
\newblock \emph{Precision electroweak measurements on the z resonance},
\newblock Physics Reports \textbf{427}(5), 257 (2006),
\newblock \doi{10.1016/j.physrep.2005.12.006}.

\bibitem{RK_21}
{R. Aaij et al. [LHCb collaboration]},
\newblock \emph{Test of lepton universality in beauty-quark decays} (2021),
  \eprint{2103.11769}.

\bibitem{LQ_14}
G.~Hiller and M.~Schmaltz,
\newblock \emph{{$R_K$ and future $b\to s\ell\ell$ physics beyond the standard
  model opportunities}},
\newblock Physical Review D \textbf{90}(5) (2014),
\newblock \doi{10.1103/physrevd.90.054014}.

\bibitem{LQ_15}
B.~Gripaios, M.~Nardecchia and S.~A. Renner,
\newblock \emph{{Composite leptoquarks and anomalies in $B$-meson decays}}
  (2015), \eprint{1412.1791}.

\bibitem{LQ_15_2}
I.~de~Medeiros~Varzielas and G.~Hiller,
\newblock \emph{Clues for flavor from rare lepton and quark decays} (2015),
  \eprint{1503.01084}.

\bibitem{LQ_16}
R.~Barbieri, C.~W. Murphy and F.~Senia,
\newblock \emph{B-decay anomalies in a composite leptoquark model},
\newblock The European Physical Journal C \textbf{77}(1) (2016),
\newblock \doi{10.1140/epjc/s10052-016-4578-7}.

\bibitem{Zprime_14}
W.~Altmannshofer, S.~Gori, M.~Pospelov and I.~Yavin,
\newblock \emph{{Quark flavor transitions in $L_\mu$ - $L_\tau$ models}},
\newblock Physical Review D \textbf{89}(9) (2014),
\newblock \doi{10.1103/physrevd.89.095033}.

\bibitem{Zprime_15}
A.~Crivellin, G.~D’Ambrosio and J.~Heeck,
\newblock \emph{{Explaining $h\to \mu^\pm \tau^\mp $, $B\to K^*\mu^+\mu^-$, and
  $B\to K \mu^+\mu^-/B\to Ke^+e^-$ in a Two-Higgs-Doublet Model with Gauged
  $L_\mu - L_\tau$}},
\newblock Physical Review Letters \textbf{114}(15) (2015),
\newblock \doi{10.1103/physrevlett.114.151801}.

\bibitem{Zprime_15_2}
A.~Celis, J.~Fuentes-Martín, M.~Jung and H.~Serôdio,
\newblock \emph{{Family nonuniversal $Z'$ models with protected flavor-changing
  interactions}},
\newblock Physical Review D \textbf{92}(1) (2015),
\newblock \doi{10.1103/physrevd.92.015007}.

\bibitem{Zprime_15_3}
A.~Falkowski, M.~Nardecchia and R.~Ziegler,
\newblock \emph{Lepton flavor non-universality in b-meson decays from a u(2)
  flavor model},
\newblock Journal of High Energy Physics \textbf{2015}(11) (2015),
\newblock \doi{10.1007/jhep11(2015)173}.

\bibitem{nu_lfv}
{B.J. Rebel et al, [MINOS collaboration]},
\newblock \emph{First minos results with the numi beam},
\newblock Nuclear Physics B - Proceedings Supplements \textbf{168}, 195 (2007),
\newblock \doi{10.1016/j.nuclphysbps.2007.02.014}.

\bibitem{LHCb_performances}
{R. Aaij et al. [LHCB collaboration]},
\newblock \emph{Lhcb detector performance},
\newblock Int. J. Mod. Phys. A \textbf{30}(07), 1530022 (2015),
\newblock \doi{10.1142/S0217751X15300227}.

\bibitem{RK_pred}
{Bordone, M., Isidori, G. \& Pattori, A},
\newblock \emph{{On the standard model predictions for $R_K$ and $R_{K^*}$}},
\newblock The European Physical Journal C \textbf{76}(8), 440 (2016),
\newblock \doi{10.1140/epjc/s10052-016-4274-7}.

\bibitem{RK_19}
{R. Aaij et al. [LHCB collaboration]},
\newblock \emph{Search for lepton-universality violation in
  ${B}^{+}\ensuremath{\rightarrow}{K}^{+}{\ensuremath{\ell}}^{+}{\ensuremath{\ell}}^{\ensuremath{-}}$
  decays},
\newblock Phys. Rev. Lett. \textbf{122}, 191801 (2019),
\newblock \doi{10.1103/PhysRevLett.122.191801}.

\bibitem{RKst}
{R. Aaij et al. [LHCB collaboration]},
\newblock \emph{{Test of lepton universality with $B^0\to K^{*0}\ell^+\ell^-$
  decays}},
\newblock Journal of High Energy Physics \textbf{2017}(8) (2017),
\newblock \doi{10.1007/jhep08(2017)055}.

\bibitem{RpK}
{R. Aaij et al. [LHCB collaboration]},
\newblock \emph{{Test of lepton universality with $ {\Lambda}_b^0\to
  {pK}^{-}{\mathrm{\ell} }^{+}{\mathrm{\ell}}^{-} $ decays}},
\newblock Journal of High Energy Physics \textbf{2020}(5) (2020),
\newblock \doi{10.1007/jhep05(2020)040}.

\bibitem{RKshort}
{R. Aaij et al. [LHCB collaboration]},
\newblock \emph{{Tests of lepton universality using $B^0\to K^0_S \ell^+
  \ell^-$ and $B^+\to K^{*+} \ell^+ \ell^-$ decays}} (2021),
  \eprint{2110.09501}.

\bibitem{angular_rate}
W.~Altmannshofer, P.~Ball, A.~Bharucha, A.~J. Buras, D.~M. Straub and M.~Wick,
\newblock \emph{{Symmetries and asymmetries of $B\to K^* \mu^+\mu^-$ decays in
  the Standard Model and beyond}},
\newblock Journal of High Energy Physics \textbf{2009}(01), 019–019 (2009),
\newblock \doi{10.1088/1126-6708/2009/01/019}.

\bibitem{Angular_Kst0}
{R. Aaij et al. [LHCB collaboration]},
\newblock \emph{{Measurement of CP-Averaged Observables in the $B^0\to
  K^{*0}\mu^+\mu^-$ Decay}},
\newblock Physical Review Letters \textbf{125}(1) (2020),
\newblock \doi{10.1103/physrevlett.125.011802}.

\bibitem{Angular_Kstplus}
{R. Aaij et al. [LHCB collaboration]},
\newblock \emph{{Angular Analysis of the $B^+\to K^{*0}\mu^+\mu^-$ Decay}},
\newblock Physical Review Letters \textbf{126}(16) (2021),
\newblock \doi{10.1103/physrevlett.126.161802}.

\bibitem{Angular_phi}
{R. Aaij et al. [LHCB collaboration]},
\newblock \emph{{Angular analysis of the rare decay $B_s^0\to\phi\mu^+\mu^-$}}
  (2021), \eprint{2107.13428}.

\bibitem{BR_phi}
{R. Aaij et al. [LHCB collaboration]},
\newblock \emph{{Branching Fraction Measurements of the Rare
  $B^0_s\to\phi\mu^+\mu^-$ and $B^0_s\to f'_2 (1525)\mu^+\mu^-$ Decays}},
\newblock Physical Review Letters \textbf{127}(15) (2021),
\newblock \doi{10.1103/physrevlett.127.151801}.

\bibitem{BR_Kst}
{R. Aaij et al. [LHCB collaboration]},
\newblock \emph{{Differential branching fractions and isospin asymmetries of
  $B\to K^(*)\mu^+\mu^-$ decays}},
\newblock Journal of High Energy Physics \textbf{2014}(6) (2014),
\newblock \doi{10.1007/jhep06(2014)133}.

\bibitem{BR_Kplus}
{R. Aaij et al. [LHCB collaboration]},
\newblock \emph{{Measurements of the S-wave fraction in $B^{0}\rightarrow
  K^{+}\pi^{-}\mu^{+}\mu^{-}$ decays and the $B^{0}\rightarrow
  K^{\ast}(892)^{0}\mu^{+}\mu^{-}$ differential branching fraction.
  Measurements of the S-wave fraction in $B^{0}\rightarrow
  K^{+}\pi^{-}\mu^{+}\mu^{-}$ decays and the $B^{0}\rightarrow
  K^{\ast}(892)^{0}\mu^{+}\mu^{-}$ differential branching fraction}},
\newblock JHEP \textbf{11}, 047. 33 p (2016),
\newblock \doi{10.1007/JHEP11(2016)047}.

\bibitem{Bmumu}
{R. Aaij et al. [LHCB collaboration]},
\newblock \emph{{Measurement of the $B^0_s\to\mu^+\mu^-$ decay properties and
  search for the $B^0\to\mu^+\mu^-$ and $B^0_s\to\mu^+\mu^-\gamma$ decays}}
  (2021),
\newblock \eprint{2108.09283}.

\bibitem{C9_C10}
{Altmannshofer, Wolfgang and Stangl, Peter},
\newblock \emph{New physics in rare b decays after moriond 2021}  (2021),
\newblock \eprint{2103.13370}.

\bibitem{BR_LHCb_Btautau}
{R. Aaij et al. [LHCB collaboration]},
\newblock \emph{{Search for the Decays
  ${B}_{s}^{0}\ensuremath{\rightarrow}{\ensuremath{\tau}}^{+}{\ensuremath{\tau}}^{\ensuremath{-}}$
  and
  ${B}^{0}\ensuremath{\rightarrow}{\ensuremath{\tau}}^{+}{\ensuremath{\tau}}^{\ensuremath{-}}$}},
\newblock Phys. Rev. Lett. \textbf{118}, 251802 (2017),
\newblock \doi{10.1103/PhysRevLett.118.251802}.

\bibitem{BR_BaBar_BKtautau}
{J. Lees, et al. [BaBar Collaboration]},
\newblock \emph{{Search for
  ${B}^{+}\ensuremath{\rightarrow}{K}^{+}{\ensuremath{\tau}}^{+}{\ensuremath{\tau}}^{\ensuremath{-}}$
  at the BaBar Experiment}},
\newblock Phys. Rev. Lett. \textbf{118}, 031802 (2017),
\newblock \doi{10.1103/PhysRevLett.118.031802}.

\bibitem{BR_bstautau}
\emph{{Searching for New Physics with
  $b\ensuremath{\rightarrow}s{\ensuremath{\tau}}^{+}{\ensuremath{\tau}}^{\ensuremath{-}}$
  Processes}},
\newblock Phys. Rev. Lett. \textbf{120}, 181802 (2018),
\newblock \doi{10.1103/PhysRevLett.120.181802}.

\bibitem{RDst_muonic}
{R. Aaij et al. [LHCB collaboration]},
\newblock \emph{{Measurement of the Ratio of Branching Fractions
  $\mathcal{B}({\overline{B}}^{0}\ensuremath{\rightarrow}{D}^{*+}{\ensuremath{\tau}}^{\ensuremath{-}}{\overline{\ensuremath{\nu}}}_{\ensuremath{\tau}})/\mathcal{B}({\overline{B}}^{0}\ensuremath{\rightarrow}{D}^{*+}{\ensuremath{\mu}}^{\ensuremath{-}}{\overline{\ensuremath{\nu}}}_{\ensuremath{\mu}})$}},
\newblock Phys. Rev. Lett. \textbf{115}, 111803 (2015),
\newblock \doi{10.1103/PhysRevLett.115.111803}.

\bibitem{RDst_hadronic}
{R. Aaij et al. [LHCB collaboration]},
\newblock \emph{{Measurement of the Ratio of the
  ${B}^{0}\ensuremath{\rightarrow}{D}^{*\ensuremath{-}}{\ensuremath{\tau}}^{+}{\ensuremath{\nu}}_{\ensuremath{\tau}}$
  and
  ${B}^{0}\ensuremath{\rightarrow}{D}^{*\ensuremath{-}}{\ensuremath{\mu}}^{+}{\ensuremath{\nu}}_{\ensuremath{\mu}}$
  Branching Fractions Using Three-Prong $\ensuremath{\tau}$-Lepton Decays}},
\newblock Phys. Rev. Lett. \textbf{120}, 171802 (2018),
\newblock \doi{10.1103/PhysRevLett.120.171802}.

\bibitem{HFLAV}
{Amhis, Yasmine Sara et al. [HFLAV Collaboration]},
\newblock \emph{{Averages of $b$-hadron, $c$-hadron, and $\tau$-lepton
  properties as of 2018}},
\newblock Eur. Phys. J. \textbf{C81}, 226 (2021),
\newblock \doi{10.1140/epjc/s10052-020-8156-7}.

\bibitem{LFV_LQ}
I.~de~Medeiros~Varzielas and G.~Hiller,
\newblock \emph{Clues for flavor from rare lepton and quark decays} (2015),
  \eprint{1503.01084}.

\bibitem{Kmutau}
{R. Aaij et al. [LHCB collaboration]},
\newblock \emph{{Search for the lepton flavour violating decay $B^+\to
  K^+\mu^-\tau^+$ using $B^{*0}_{s2}$ decays}},
\newblock Journal of High Energy Physics (6), 129 (2020),
\newblock \doi{10.1007/JHEP06(2020)129}.

\bibitem{Kmue}
{R. Aaij et al. [LHCB collaboration]},
\newblock \emph{{Search for Lepton-Flavor Violating Decays
  ${B}^{+}\ensuremath{\rightarrow}{K}^{+}{\ensuremath{\mu}}^{\ifmmode\pm\else\textpm\fi{}}{e}^{\ensuremath{\mp}}$}},
\newblock Phys. Rev. Lett. \textbf{123}, 241802 (2019),
\newblock \doi{10.1103/PhysRevLett.123.241802}.

\bibitem{mutau}
{R. Aaij et al. [LHCB collaboration]},
\newblock \emph{{Search for the Lepton-Flavor-Violating Decays
  ${B}_{s}^{0}\ensuremath{\rightarrow}{\ensuremath{\tau}}^{\ifmmode\pm\else\textpm\fi{}}{\ensuremath{\mu}}^{\ensuremath{\mp}}$
  and
  ${B}^{0}\ensuremath{\rightarrow}{\ensuremath{\tau}}^{\ifmmode\pm\else\textpm\fi{}}{\ensuremath{\mu}}^{\ensuremath{\mp}}$}},
\newblock Phys. Rev. Lett. \textbf{123}, 211801 (2019),
\newblock \doi{10.1103/PhysRevLett.123.211801}.

\bibitem{mue}
{R. Aaij et al. [LHCB collaboration]},
\newblock \emph{{Search for the lepton-flavour violating decays $B_{(s)}^0 \to
  e^\pm \mu^\mp$}},
\newblock Journal of High Energy Physics \textbf{2018}(3), 78 (2018),
\newblock \doi{10.1007/JHEP03(2018)078}.

\bibitem{Dhll}
{R. Aaij et al. [LHCB collaboration]},
\newblock \emph{{Searches for 25 rare and forbidden decays of $D^+$ and $D^+_s$
  mesons}},
\newblock Journal of High Energy Physics \textbf{2021}(6), 44 (2021),
\newblock \doi{10.1007/JHEP06(2021)044}.

\end{thebibliography}

\nolinenumbers

\end{document}